# Adaptive Multicarrier Quadrature Division Modulation for Continuous-Variable Quantum Key Distribution

Laszlo Gyongyosi

[1] Quantum Technologies Laboratory, Department of Telecommunications
*Budapest University of Technology and Economics*
2 Magyar tudosok krt, Budapest, *H*-1117, Hungary
[2] Information Systems Research Group, Mathematics and Natural Sciences
*Hungarian Academy of Sciences*
Budapest, *H*-1518, Hungary

gyongyosi@hit.bme.hu

**Abstract**

In a continuous-variable quantum key distribution (CVQKD) system, the information is conveyed by coherent state carriers. The quantum continuous variables are sent through a quantum channel, where the presence of the eavesdropper adds a white Gaussian noise to the transmission. The amount of tolerable noise and loss is a crucial point in CVQKD, since it determines the overall performance of the protocol, including the secure key rates and transmission distances. In this work, we propose the adaptive multicarrier quadrature division (AMQD) modulation technique for CVQKD. The method granulates the Gaussian random input into Gaussian subcarrier continuous variables in the encoding phase, which are then decoded by a continuous unitary transformation. The subcarrier coherent variables formulate Gaussian sub-channels from the physical link with strongly diverse transmission capabilities, which leads to significantly improved transmission efficiency, higher tolerable loss, and excess noise. We also investigate a modulation-variance adaption technique within the AMQD scheme, which provides optimal capacity-achieving communication over the sub-channels in the presence of a Gaussian noise.

# 1 Introduction

Continuous-variable (CV) quantum key distribution (QKD) systems allow for the establishment of an unconditionally secure quantum communication over the current standard telecommunication networks. CVQKD systems possess several benefits and advantages over the DV (discrete variable) protocols, since they do not require specialized devices or unreachable special requirements in an experimental scenario [11-18]. CVQKD systems are based on continuous variables such as Gaussian random position and momentum quadratures in the phase space. The Gaussian modulated coherent states are transmitted over a noisy quantum channel, where the presence of an eavesdropper adds a white Gaussian noise to the transmission. Since CVQKD schemes were developed and introduced just a few years ago, there are still many open questions regarding the optimal encoding scheme. A Gaussian modulation is a robust and easily applicable finding in a practical scenario, and allows for the implementation of the protocol in the experiment; however, CVQKD is still very sensitive to the imperfections of the transmission and the practical devices. The performance of the protocol is strongly determined by the excess noise of the quantum channel, and the transmittance parameter of the physical link (specifically, the Gaussian noise of the quantum channel models the eavesdropper's optimal entangling cloner attack [2-3], [11-18], and the channel is referred to as a Gaussian quantum channel). Since the amount of tolerable loss and the excess noise are central parameters from the viewpoint of the running of CVQKD, it would be very desirable to make some optimization steps in the encoding and decoding process to override the current limitations and to improve the quality of the quantum-level transmission. Our aim is to provide a solution to this problem by introducing the adaptive multicarrier quadrature division (AMQD) modulation technique for CVQKD, which can be applied both in one-way and two-way CVQKD to increase the tolerable loss and excess noise. In traditional telecommunications, OFDM (orthogonal frequency-division multiplexing) is a well-known and widely applied technique for improving the bandwidth efficiency over noisy communication networks [6-10]. In an OFDM scheme, the information is encoded in multiple carrier frequencies, and its main advantage over single-carrier transmission is that the subcarrier-based transmission can attenuate and overwhelm the problems of diverse and unfavorable channel conditions. OFDM systems have been admitted to be a useful encoding method in traditional networking; however, no similar method exists for CVQKD. If a similar solution were available for continuous variables, one could enjoy similar benefits in a quantum-communication scenario; however, up to this point no analogous answer exists for quantum-level transmission. With this in mind, we introduce the idea of AMQD, which works on continuous variables and for which similar benefits can be reached in the process of quantum-level information transmission, e.g., in a classical scenario by the application of the OFDM. In the standard coding scenario, Alice, the sender, modulates and separately transmits each coherent state in the phase space. This standard modulation scheme is referred as *single-carrier* modulation throughout, consistent to its traditional meaning.

The key idea behind AMQD modulation is as follows. Alice, draws a zero-mean, circular symmetric complex Gaussian random vector, which is then transformed by the inverse Fourier operation. At a given modulation variance, Alice prepares her *Gaussian subcarrier* CVs, which are then fed into the. Bob, the receiver, applies the inverse unitary of Alice's operation, which makes it possible for him to recover the noisy version of Alice's input coherent states. This kind of communication will be referred as *multicarrier* modulation.

What are the main advantages of this kind of communication? There are several fine corollaries. First of all, the Gaussian subcarrier CV states sent through the channel, which overall allows higher tolerable loss and excess noise at a given modulation variance. Second, the Gaussian quantum channel can be viewed as several parallel Gaussian quantum channels, called *sub-channels*, each dedicated for the transmission of a given subcarrier with an independent, and significantly lower noise variance. Third, the information transmission capability of the sub-channels is very diverse, depending on the variance of the subcarrier CV, which allows for the development of smart adaptive modulation techniques for the proposed multicarrier quadrature division-encoding technique. The idea behind this is to use only the "good" Gaussian sub-channels for the transmission, and to not send any valuable information over the so noisy sub-channels. It is a particularly convenient approach, since the result of the adaptive allocation is a better performance of the protocol at low SNRs (signal-to-noise ratio) and higher tolerable loss, which are crucial cornerstones in an experimental CVQKD that operates in practice at very low SNRs. From these, the purposes are now clear. We have to find the operation that works on continuous variables and outputs the subcarrier quadratures, which can divide the physical Gaussian channel into Gaussian sub-channels. We also need the continuous unitary inverse of this operation. If we have it, then we have to find an adaptive modulation-variance allocation mechanism, which allows no to send valuable information over the very noisy Gaussian sub-channels. Fortunately, these features are all included in our AMQD coding scheme. The AMQD modulation granulates Alice's initial Gaussian states into several subcarrier Gaussian CVs, which divide the physical channel into several Gaussian sub-channels. Bob applies an inverse continuous unitary operation, which allows him to obtain Alice's initial (noisy) coherent states. The proposed AMQD modulation offers several important features, but the main improvement is in the quality of the quantum-level transmission, since the subcarriers allow a more efficient communication over the same quantum channel at a given modulation variance.

This paper is organized as follows. In Section 2, preliminaries are proposed. Section 3 introduces the multicarrier quadrature division scheme. In Section 4, the adaptive modulation variance allocation mechanism is discussed. Section 5 studies the performance of AMQD modulation. Section 6 provides the security proof of AMQD against optimal Gaussian collective attacks. Finally, Section 7 concludes the results.

## 2 Preliminaries

In the standard single-carrier modulation scheme, the input coherent state $|\varphi_i\rangle = |x_i + \mathrm{i}p_i\rangle$ is a Gaussian state in the phase space $\mathcal{S}$, with i.i.d. Gaussian random position and momentum quadratures $x_i \in \mathbb{N}\left(0, \sigma_{\omega_0}^2\right)$, $p_i \in \mathbb{N}\left(0, \sigma_{\omega_0}^2\right)$, where $\sigma_{\omega_0}^2$ is the modulation variance. The coherent state $|\varphi_i\rangle$ in the phase space $\mathcal{S}$ can be modeled as a zero-mean, circular symmetric complex Gaussian random variable $z \in \mathcal{CN}\left(0, \sigma_{\omega_z}^2\right)$, with variance $\sigma_{\omega_z}^2 = \mathbb{E}\left[|z|^2\right]$, and with i.i.d. real and

imaginary zero-mean Gaussian random components, $\operatorname{Re}(z_i) \in \mathbb{N}(0, \sigma_{\omega_0}^2)$, $\operatorname{Im}(z_i) \in \mathbb{N}(0, \sigma_{\omega_0}^2)$.

In the single-carrier scenario, the transmission of this complex variable over the Gaussian quantum channel $\mathcal{N}$ can be characterized by the $T(\mathcal{N})$ normalized complex transmittance variable

$$T(\mathcal{N}) = \operatorname{Re} T(\mathcal{N}) + \mathrm{i} \operatorname{Im} T(\mathcal{N}) \in \mathcal{C}, \tag{1}$$

where $0 \leq \operatorname{Re} T(\mathcal{N}) \leq 1/\sqrt{2}$ stands for the transmission of the position quadrature, $0 \leq \operatorname{Im} T(\mathcal{N}) \leq 1/\sqrt{2}$ is the transmission of the momentum quadrature, with relation

$$\operatorname{Re} T(\mathcal{N}) = \operatorname{Im} T(\mathcal{N}) \tag{2}$$

by our convention. The $0 \leq |T(\mathcal{N})| \leq 1$ magnitude of the $T(\mathcal{N})$ complex variable is

$$|T(\mathcal{N})| = \sqrt{\operatorname{Re} T(\mathcal{N})^2 + \operatorname{Im} T(\mathcal{N})^2} = \sqrt{2} \operatorname{Re} T(\mathcal{N}) \in \mathbb{R} \tag{3}$$

and the squared magnitude of $T(\mathcal{N})$ is

$$|T(\mathcal{N})|^2 = \operatorname{Re} T(\mathcal{N})^2 + \operatorname{Im} T(\mathcal{N})^2 = 2 \operatorname{Re} T(\mathcal{N})^2 \in \mathbb{R}, \tag{4}$$

where $0 \leq |T(\mathcal{N})|^2 \leq 1$. Assuming a 0 dB loss, the quadrature transmittance is parameterized with $\operatorname{Re} T(\mathcal{N}) = \operatorname{Im} T(\mathcal{N}) = \frac{1}{\sqrt{2}}$ and

$$|T(\mathcal{N})| = |T(\mathcal{N})|^2 = 1. \tag{5}$$

At a given input $x$, the channel output $y$ can be expressed as

$$y = T(\mathcal{N})x + \Delta, \tag{6}$$

where $\Delta \in \mathcal{CN}(0, \sigma_\Delta^2)$, $\sigma_\Delta^2 = \mathbb{E}\left[|\Delta|^2\right]$, models the Gaussian noise of the quantum channel (also a zero-mean, symmetric circular complex Gaussian random variable), with independent quadrature components $\Delta_x \in \mathbb{N}(0, \sigma_\mathcal{N}^2)$, $\Delta_p \in \mathbb{N}(0, \sigma_\mathcal{N}^2)$.

This Gaussian quantum channel is equipped with a capacity of

$$C(\mathcal{N}) = \log_2\left(1 + \frac{\sigma_{\omega_0}^2 |T(\mathcal{N})|^2}{\sigma_\mathcal{N}^2}\right), \tag{7}$$

where $\sigma_{\omega_0}^2$ is the modulation variance (single-carrier) of the input Gaussian signal.

In the multicarrier case, the Gaussian quantum channel is divided into $n$ Gaussian sub-channels $\mathcal{N}_i$, $i = 1 \ldots n$, each with an independent noise variance $\sigma_{\mathcal{N}_i}^2$, each for the transmission of a continuous variable subcarrier $|\phi_i\rangle$, which leads to the output for the $i$-th sub-channel:

$$y_i = T(\mathcal{N}_i)x_i + \Delta_i, \ i = 1\ldots n, \tag{8}$$

where $T(\mathcal{N}_i) \in \mathcal{C}$, $\Delta_i \in \mathcal{CN}(0, \sigma^2_{\Delta_i})$, and the resulting transmission capacity is

$$C(\mathcal{N}) = \max_{\forall i} \sum_{i=1}^{n} \log_2 \left(1 + \frac{\sigma^2_\omega |T(\mathcal{N}_i)|^2}{\sigma^2_{\mathcal{N}_i}}\right), \tag{9}$$

where $\sigma^2_\omega = \frac{1}{n}\sum_{i=1}^{n} \sigma^2_{\omega_i} = \sigma^2_{\omega_0}$ is the average modulation variance of the $i$-th Gaussian subcarrier CV transmitted via the $i$-th Gaussian sub-channel.

At this point, the capacity formulas of (7) and (9) require some clarification. Assuming a Gaussian quantum channel, one can find two different capacity formulas for the real dimension and the complex dimension. The reason is as follows. The noise is independent on the real and imaginary parts (i.e., on the position and momentum quadratures), and each use of the complex Gaussian channel is in particular analogous to two uses of the real Gaussian channel. From this distinction, two different types of capacity formulas can be derived for the same Gaussian channel, i.e., the *real-dimension* capacity and the *complex-dimension* capacity.

The real dimension capacity of an AWGN is

$$C(\mathcal{N}) = \tfrac{1}{2}\log_2\left(1 + \frac{\sigma^2_{\omega_0}|T(\mathcal{N})|^2}{\sigma^2_\mathcal{N}}\right), \tag{10}$$

while the complex dimension capacity is precisely

$$C(\mathcal{N}) = \log_2\left(1 + \frac{\sigma^2_\omega |T(\mathcal{N})|^2}{\sigma^2_\mathcal{N}}\right). \tag{11}$$

In (11) the use of the complex domain is justified by the fact that circular symmetric complex Gaussian random variables will be transmitted through the Gaussian quantum channel.

The SNR of the channel is

$$\text{SNR} = \frac{\sigma^2_{\omega_0}|T(\mathcal{N})|^2}{\sigma^2_\mathcal{N}}. \tag{12}$$

For the precise clarification, we will use the complex-domain capacity formula throughout, since the coherent state in the phase space is a complex variable, despite the fact that the quadrature measurement process will finally lead to the formula of (10).

## 3 Multicarrier Quadrature Division Modulation

### 3.1 Continuous-Variable Quantum Fourier Transform

In terms of the CV scenario, by a convention the $|x\rangle$ position quadrature could be used as a computational basis. We will also do this throughout. (The continuous-variable quantum Fourier transformation will be abbreviated as CVQFT.)

Let the Gaussian variable be

$$g(x) = \frac{1}{\sigma\sqrt{2\pi}} e^{\frac{-x^2}{2\sigma^2}}, \ x \in \mathbb{N}(0, \sigma^2), \tag{13}$$

where $\int_{-\infty}^{\infty} g(x) dx = 1$. The Fourier transform of (13) is expressed as

$$F(g(x)) = G(\omega) = e^{\frac{-\omega^2 \sigma^2}{2}}, \ \omega \in \mathbb{N}(0, \sigma^2). \tag{14}$$

Between the Fourier transform $F(x)$ and the inverse Fourier transform function $F^{-1}(\omega)$, the connection is as follows [7-8]:

$$F(x) = \int_{-\infty}^{\infty} F^{-1}(\omega) e^{-\mathrm{i}x\omega} d\omega, \tag{15}$$

where

$$F^{-1}(\omega) = \frac{1}{2\pi} \int_{-\infty}^{\infty} F(x) e^{\mathrm{i}x\omega} dx. \tag{16}$$

For an $n$-dimensional Gaussian random vector $\mathbf{g} \in (0, \mathbf{K_g})$, $\mathbf{g} = (g_1,, g_n)^T$, where $\mathbf{K_g} = \mathbb{E}[\mathbf{gg}^T]$ is the covariance matrix, the Fourier transform and its inverse:

$$F(\mathbf{g}) = \int_{-\infty}^{\infty} \int_{-\infty}^{\infty} \ldots \int_{-\infty}^{\infty} F^{-1}(\mathbf{w}) e^{-\mathrm{i}\mathbf{g}\cdot\mathbf{w}} d\mathbf{w}, \tag{17}$$

where $\mathbf{w} \in (0, \mathbf{K_w})$ is a $n$-dimensional Gaussian random vector with covariance matrix $\mathbf{K_w} = \mathbb{E}[\mathbf{ww}^T]$, and

$$F^{-1}(\mathbf{w}) = \frac{1}{(2\pi)^n} \int_{-\infty}^{\infty} \int_{-\infty}^{\infty} \ldots \int_{-\infty}^{\infty} F(\mathbf{g}) e^{\mathrm{i}\mathbf{g}\cdot\mathbf{w}} d\mathbf{g}. \tag{18}$$

**Proposition 1.** *The $\sigma_F^2$ variance of a Gaussian subcarrier CV $|\phi_i\rangle$ is the reciprocal of $\sigma_{\omega_0}^2$, $\sigma_F^2 = \frac{1}{\sigma_{\omega_0}^2}$, where $\sigma_{\omega_0}^2$ is the single-carrier modulation variance.*

*Proof.*
First, we propose the CVQFT operation acting on the continuous variables, and then we rewrite it as a zero-mean, circular symmetric complex Gaussian random variable.
Assuming $|x\rangle$ position computational basis, function $F(\cdot)$ acts on the coherent input $|\varphi_i\rangle$ as follows [1]:

$$F(|\varphi_i\rangle) = F(\langle x|\varphi_i\rangle) \to \langle p|\varphi_i\rangle, \tag{19}$$

where

$$\langle x|\varphi_i\rangle = \frac{1}{\sqrt{2\pi}} \int_{-\infty}^{\infty} \langle p|\varphi_i\rangle dp e^{ipx}, \qquad (20)$$

and

$$\langle p|\varphi_i\rangle = \frac{1}{\sqrt{2\pi}} \int_{-\infty}^{\infty} \langle x|\varphi_i\rangle dx e^{-ipx} \qquad (21)$$

wavefunctions in the position space $x$, and momentum space $p$ for the state $|\varphi_i\rangle$, respectively. The notation $\langle x|\psi\rangle$ stands for the inner product [1], which can be rewritten as

$$\langle x|\psi\rangle = \int_{-\infty}^{\infty} f(x-x')\psi(x')dx', \qquad (22)$$

where $f(x-x') = \langle x|x'\rangle$ and $\psi(x) = \langle x|\psi\rangle$. Assuming a complete set of orthonormal wavefunctions $\{\varphi_n\}$ [1], the arbitrary wavefunction $\psi$ is as $\psi = \sum_n c_i \varphi_i$, from which

$$\psi(q) = \int_{-\infty}^{\infty} c(p)u(p;q)dp, \qquad (23)$$

where

$$\begin{aligned}
c(p) &= \langle u(p;q), \psi(q)\rangle \\
&= \left\langle u(p;q), \int_{-\infty}^{\infty} c(p)u(p;q)dp \right\rangle \\
&= \int_{-\infty}^{\infty} f(p-p')c(p')dp'.
\end{aligned} \qquad (24)$$

The inverse function of (19) is defined as

$$F^{-1}(\langle p|\varphi_i\rangle) = \langle x|\varphi_i\rangle. \qquad (25)$$

A Gaussian modulated coherent state $|\varphi_i\rangle = |x_i + ip_i\rangle$, where $x_i \in \mathbb{N}(0, \sigma_{\omega_0}^2)$, $p_i \in \mathbb{N}(0, \sigma_{\omega_0}^2)$ are the position and momentum quadratures, respectively; can be rewritten as a zero-mean, circular symmetric complex Gaussian random variable $z_i \in \mathcal{CN}(0, \sigma_{\omega_{z_i}}^2)$, $\sigma_{\omega_{z_i}}^2 = \left[\mathbb{E}|z_i|^2\right]$, as

$$z_i = x_i + ip_i, \qquad (26)$$

where $x_i \in \mathbb{N}(0, \sigma_{\omega_0}^2)$, $p_i \in \mathbb{N}(0, \sigma_{\omega_0}^2)$ are i.i.d. zero-mean Gaussian random variables. As follows,

$$|\varphi_i\rangle = |z_i\rangle. \qquad (27)$$

The variable $e^{i\varphi_i} z_i$ has the same distribution of $z_i$ for any $\varphi_i$, i.e., $\mathbb{E}[z_i] = \mathbb{E}[e^{i\varphi_i} z_i] = \mathbb{E}e^{i\varphi_i}[z_i]$ and $\sigma_{z_i}^2 = \mathbb{E}[|z_i|^2]$. The density of $z_i$ is

$$f(z_i) = \frac{1}{2\pi\sigma_{\omega_0}^2} e^{\frac{-(|z_i|^2)}{2\sigma_{\omega_0}^2}} = f(x_i, p_i) = \frac{1}{2\pi\sigma_{\omega_0}^2} e^{\frac{-(x_i^2 + p_i^2)}{2\sigma_{\omega_0}^2}}, \tag{28}$$

where $|z_i| = \sqrt{x_i^2 + p_i^2}$ is the magnitude, which is a Rayleigh random variable with density

$$f(|z_i|) = \frac{|z_i|}{\sigma_{\omega_{z_i}}^2} e^{\frac{-|z_i|^2}{2\sigma_{\omega_{z_i}}^2}}, |z_i| \geq 0, \tag{29}$$

while the $|z_i|^2 = x_i^2 + p_i^2$ squared magnitude is exponentially distributed with density

$$f(|z_i|^2) = \frac{1}{\sigma_{\omega_{z_i}}^2} e^{\frac{-|z_i|^2}{\sigma_{\omega_{z_i}}^2}}, |z_i|^2 \geq 0. \tag{30}$$

The $i$-th *subcarrier* CV is defined as

$$|\phi_i\rangle = |\text{IFFT}(z_i)\rangle = |F^{-1}(z_i)\rangle = |d_i\rangle, \tag{31}$$

where IFFT stands for the Inverse Fast Fourier Transform, and subcarrier continuous variable $|\phi_i\rangle$ in (31) is also a zero-mean, circular symmetric complex Gaussian random variable $d_i \in \mathcal{CN}(0, \sigma_{d_i}^2)$, $\sigma_{d_i}^2 = \mathbb{E}[|d_i|^2]$, $d_i = x_{d_i} + i p_{d_i}$, where $x_{d_i} \in \mathbb{N}(0, \sigma_{\omega_F}^2)$, $p_{d_i} \in \mathbb{N}(0, \sigma_{\omega_F}^2)$ are i.i.d. zero-mean Gaussian random variables, and $\sigma_{\omega_F}^2$ is the variance of the Fourier transformed Gaussian signal.

The inverse of (31) results the single-carrier CV from the subcarrier CV as follows:

$$|\varphi_i\rangle = \text{CVQFT}(|\phi_i\rangle) = F(|d_i\rangle) = |F(F^{-1}(z_i))\rangle = |z_i\rangle, \tag{32}$$

where CVQFT is the continuous-variable QFT operation.

Let us to derive the Fourier transform of the Gaussian input [7-8]. First, we rewrite (28) in the position basis $x$ as $g(x) = \frac{1}{\sqrt{2\pi\sigma_{\omega_0}^2}} e^{\frac{-x^2}{a^2}}$, where $a^2 = 2\sigma_{\omega_0}^2$. Then, the Fourier transformed signal $G(p)$ is precisely evaluated as

$$\begin{aligned}
F\big(g(x)\big) &= G(p) \\
&= \frac{1}{\sqrt{2\pi}} \frac{1}{\sqrt{2\pi\sigma_{\omega_0}^2}} \int_{-\infty}^{\infty} g(x) e^{-\mathrm{i}px} dx \\
&= \frac{1}{\sqrt{2\pi}} \frac{1}{\sqrt{2\pi\sigma_{\omega_0}^2}} \int_{-\infty}^{\infty} e^{\frac{-x^2}{a^2}} e^{-\mathrm{i}px} dx \\
&= \frac{1}{\sqrt{2\pi}} \frac{1}{\sqrt{2\pi\sigma_{\omega_0}^2}} \int_{-\infty}^{\infty} e^{\left(\frac{-x^2}{a^2} - \mathrm{i}px\right)} dx \\
&= \frac{1}{\sqrt{2\pi}} \frac{1}{\sqrt{2\pi\sigma_{\omega_0}^2}} e^{-\frac{a^2 p^2}{4}} \int_{-\infty}^{\infty} e^{-\left(\frac{x}{a} + \frac{\mathrm{i}ap}{2}\right)^2} dx. \\
&= \frac{1}{\sqrt{2\pi}} \frac{1}{\sqrt{2\pi\sigma_{\omega_0}^2}} e^{\frac{-a^2 p^2}{4}} \int_{-\infty}^{\infty} e^{-\frac{\left(x + \frac{\mathrm{i}a^2 p}{2}\right)^2}{a^2}} dx \\
&= \frac{1}{\sqrt{2\pi}} \frac{1}{\sqrt{2\pi\sigma_{\omega_0}^2}} e^{-\frac{\sigma_{\omega_0}^2 p^2}{2}} = \frac{1}{\sqrt{2\pi}} \frac{1}{\sqrt{2\pi\sigma_{\omega_0}^2}} e^{-\frac{p^2}{2\sigma_F^2}} \\
&= \frac{1}{\sqrt{2\pi}} \frac{1}{\sqrt{\frac{2\pi}{\sigma_F^2}}} e^{-\frac{p^2}{2\sigma_F^2}}.
\end{aligned} \qquad (33)$$

The $G(p)$ Fourier transform of the Gaussian signal $g(x)$ is also Gaussian in the conjugate-variable space, with variance $\sigma_F^2 = \frac{1}{\sigma_{\omega_0}^2}$. In other words, the $F(|\varphi_i\rangle)$ Fourier transform of the Gaussian coherent state $|\varphi_i\rangle$ is also a Gaussian state. Since the position and momentum quadratures are Fourier-transform pairs, it follows that as the modulation variance $\sigma_\omega^2$ of the input Gaussian signal increases, the variance $\sigma_F^2$ of the Fourier transformed signal decreases. From the uncertainty principle, it can be concluded that if $\Delta x = \sigma_{\omega_0}$ (i.e., the uncertainty of the Gaussian is proportional to the standard deviation) and $\Delta p = \frac{1}{\sigma_{\omega_0}}$, then $\Delta x \Delta p = 1$.

∎

The normalized Gaussian functions $g(x) = \frac{1}{\sqrt{2\pi\sigma_{\omega_i}^2} \sqrt{2\sigma_{\omega_i}}} e^{\frac{-x^2}{2\sigma_{\omega_i}^2}}$ and $G(p) = \frac{\sqrt{2}\sigma_{\omega_i}}{\sqrt{2\pi\sigma_{\omega_i}^2}} e^{-\frac{p^2}{2\sigma_{F_i}^2}}$ are shown in Fig. 1. In terms of the quadratures, the momentum $p$ is the Fourier transform of the position $x$, and the position $x$ is the inverse-Fourier transform of the momentum $p$.

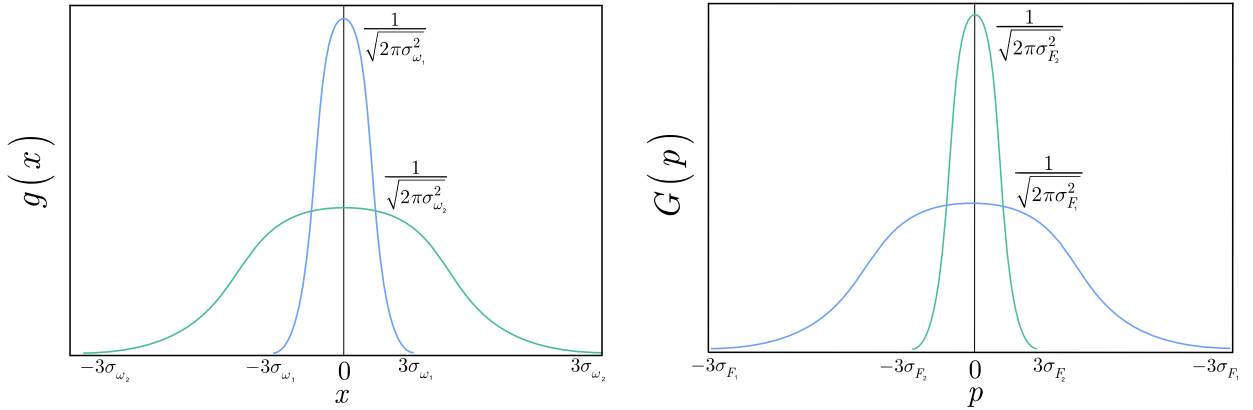

**Figure 1.** The input Gaussian state in the position basis (a) and the CVQFT transformed signal in the momentum basis, where $\sigma_{F_i}^2 = 1/\sigma_{\omega_i}^2$ (b).

The characterization of the Gaussian sub-channels is proposed Theorem 1.

**Theorem 1.** *The AMQD divides the Gaussian channel $\mathcal{N}$ into $n$ Gaussian sub-channels $\mathcal{N}_i$, $i = 1,\ldots,n$, with independent noise variances $\sigma_{\mathcal{N}_i}^2$, where $\frac{1}{n}\sum_{i=1}^n \sigma_{\mathcal{N}_i}^2 = \sigma_{\mathcal{N}}^2$.*

*Proof.*
Let $n$ is the number of Alice's input Gaussian states. The $n$ input coherent states are modeled by an $n$-dimensional, zero-mean, circular symmetric complex random Gaussian vector

$$\mathbf{z} = \mathbf{x} + \mathrm{i}\mathbf{p} = (z_1,\ldots,z_n)^T \in \mathcal{CN}(0,\mathbf{K_z}), \tag{34}$$

where each $z_i$ can be modeled as a zero-mean, circular symmetric complex Gaussian random variable $z_i \in \mathcal{CN}\left(0, \sigma_{\omega_{z_i}}^2\right)$, $z_i = x_i + \mathrm{i} p_i$.

The real and imaginary variables (i.e., the position and momentum quadratures) formulate $n$-dimensional real Gaussian random vectors, $\mathbf{x} = (x_1,\ldots,x_n)^T$ and $\mathbf{p} = (p_1,\ldots,p_n)^T$, with zero-mean Gaussian random variables

$$x_i = \frac{1}{\sigma_{\omega_0}\sqrt{2\pi}} e^{\frac{-x_i^2}{2\sigma_{\omega_0}^2}}, \; p_i = \frac{1}{\sigma_{\omega_0}\sqrt{2\pi}} e^{\frac{-p_i^2}{2\sigma_{\omega_0}^2}}, \tag{35}$$

where $\sigma_{\omega_0}^2$ is the stands for single-carrier modulation variance (precisely, the variance of the real and imaginary components of $z_i$), while $\mathbf{K_z}$ is the $n \times n$ Hermitian covariance matrix of $\mathbf{z}$:

$$\mathbf{K_z} = \mathbb{E}\left[\mathbf{z}\mathbf{z}^\dagger\right], \tag{36}$$

where $\mathbf{z}^\dagger$ is the adjoint of $\mathbf{z}$. For vector $\mathbf{z}$,

$$\mathbb{E}[\mathbf{z}] = \mathbb{E}\left[e^{\mathrm{i}\gamma}\mathbf{z}\right] = \mathbb{E}e^{\mathrm{i}\gamma}[\mathbf{z}] \tag{37}$$

holds, and

$$\mathbb{E}\left[\mathbf{z}\mathbf{z}^T\right] = \mathbb{E}\left[e^{\mathrm{i}\gamma}\mathbf{z}\left(e^{\mathrm{i}\gamma}\mathbf{z}\right)^T\right] = \mathbb{E}e^{\mathrm{i}2\gamma}\left[\mathbf{z}\mathbf{z}^T\right], \tag{38}$$

for any $\gamma \in [0, 2\pi]$. The density of $\mathbf{z}$ is as follows (if $\mathbf{K}_\mathbf{z}$ is invertible):

$$f(\mathbf{z}) = \frac{1}{\pi^n \det \mathbf{K}_\mathbf{z}} e^{-\mathbf{z}^\dagger \mathbf{K}_\mathbf{z}^{-1} \mathbf{z}}. \tag{39}$$

A $n$-dimensional Gaussian random vector is expressed as $\mathbf{x} = \mathbf{A}\mathbf{s}$, where $\mathbf{A}$ is an (invertible) linear transform from $\mathbb{R}^n$ to $\mathbb{R}^n$, and $\mathbf{s}$ is an $n$-dimensional standard Gaussian random vector $\mathbb{N}(0,1)_n$. This vector is characterized by its covariance matrix $\mathfrak{C}(\mathbf{x}) = \mathbb{E}\left[\mathbf{x}\mathbf{x}^T\right] = \mathbf{A}\mathbf{A}^T$, as

$$\mathbf{x} = \frac{1}{\left(\sqrt{2\pi}\right)^n \sqrt{\det(\mathbf{A}\mathbf{A}^T)}} e^{-\frac{\mathbf{x}^T \mathbf{x}}{2(\mathbf{A}\mathbf{A}^T)}}. \tag{40}$$

The Fourier transformation $F(\cdot)$ of the $n$-dimensional Gaussian random vector $\mathbf{v} = (v_1, \ldots, v_n)^T$ results in the $n$-dimensional Gaussian random vector $\mathbf{m} = (m_1, \ldots, m_n)^T$:

$$\mathbf{m} = F(\mathbf{v}) = e^{\frac{-\mathbf{m}^T \mathbf{A}\mathbf{A}^T \mathbf{m}}{2}} = e^{\frac{-\sigma_{\omega_0}^2 (m_1^2 + \ldots + m_n^2)}{2}}. \tag{41}$$

In the first step of AMQD, Alice applies the inverse FFT operation to vector $\mathbf{z}$ (see (34)), which results in an $n$-dimensional zero-mean, circular symmetric complex Gaussian random vector $\mathbf{d}$, $\mathbf{d} \in \mathcal{CN}(0, \mathbf{K_d})$, $\mathbf{d} = (d_1, \ldots, d_n)^T$, precisely as

$$\mathbf{d} = F^{-1}(\mathbf{z}) = e^{\frac{\mathbf{d}^T \mathbf{A}\mathbf{A}^T \mathbf{d}}{2}} = e^{\frac{\sigma_{\omega_0}^2 (d_1^2 + \ldots + d_n^2)}{2}}, \tag{42}$$

where $d_i = x_{d_i} + \mathrm{i} p_{d_i}$, $d_i \in \mathcal{CN}(0, \sigma_{d_i}^2)$, and the position and momentum quadratures of $|\phi_i\rangle$ are i.i.d. Gaussian random variables

$$x_{d_i} \in \mathbb{N}(0, \sigma_F^2), \ p_{d_i} \in \mathbb{N}(0, \sigma_F^2), \tag{43}$$

where $\mathbf{K_d} = \mathbb{E}\left[\mathbf{d}\mathbf{d}^\dagger\right]$, $\mathbb{E}[\mathbf{d}] = \mathbb{E}\left[e^{\mathrm{i}\gamma}\mathbf{d}\right] = \mathbb{E}e^{\mathrm{i}\gamma}[\mathbf{d}]$, and $\mathbb{E}\left[\mathbf{d}\mathbf{d}^T\right] = \mathbb{E}\left[e^{\mathrm{i}\gamma}\mathbf{d}\left(e^{\mathrm{i}\gamma}\mathbf{d}\right)^T\right] = \mathbb{E}e^{\mathrm{i}2\gamma}\left[\mathbf{d}\mathbf{d}^T\right]$ for any $\gamma \in [0, 2\pi]$.

In the next step, Alice modulates the coherent Gaussian subcarriers as follows:

$$|\phi_i\rangle = |d_i\rangle = |F^{-1}(z)\rangle. \tag{44}$$

The result of (42) defines $n$, independent $\mathcal{N}_i$ Gaussian sub-channels, each with noise variance

$\sigma_{\mathcal{N}_i}^2$, one for each subcarrier coherent state $|\phi_i\rangle$. After the CV subcarriers are transmitted through the noisy channel, Bob applies the CVQFT, which results him the noisy version $|\varphi_i'\rangle = |z_i'\rangle$ of Alice's input $z_i$.

On Bob's side, the received system $\mathbf{y}$ is an $n$-dimensional zero-mean, circular symmetric complex Gaussian random vector $\mathbf{y} \in \mathcal{CN}(0, \mathbb{E}[\mathbf{y}\mathbf{y}^\dagger])$. The $m$-th element of vector $\mathbf{y}$ is $y_m$, expressed as follows:

$$\begin{aligned} y_m &= F(\mathbf{T}(\mathcal{N}))z_m + F(\Delta) \\ &= F(\mathbf{T}(\mathcal{N}))F(F^{-1}(z_m)) + F(\Delta) \\ &= \sum_n F(T_i(\mathcal{N}_i))F(d_i) + F(\Delta_i), \end{aligned} \qquad (45)$$

where

$$\mathbf{T}(\mathcal{N}) = [T_1(\mathcal{N}_1), \ldots, T_n(\mathcal{N}_n)]^T \in \mathcal{C}^n, \qquad (46)$$

where

$$T_i(\mathcal{N}_i) = \operatorname{Re}(T_i(\mathcal{N}_i)) + \mathrm{i}\operatorname{Im}(T_i(\mathcal{N}_i)) \in \mathcal{C}, \qquad (47)$$

is a complex variable, which quantifies the position and momentum quadrature transmission (i.e., gain) of the $i$-th Gaussian sub-channel $\mathcal{N}_i$, in the phase space $\mathcal{S}$, with real and imaginary parts $0 \leq \operatorname{Re} T_i(\mathcal{N}_i) \leq 1/\sqrt{2}$, $0 \leq \operatorname{Im} T_i(\mathcal{N}_i) \leq 1/\sqrt{2}$. The $T_i(\mathcal{N}_i)$ variable has a magnitude of $|T_i(\mathcal{N}_i)| = \sqrt{\operatorname{Re} T_i(\mathcal{N}_i)^2 + \operatorname{Im} T_i(\mathcal{N}_i)^2} \in \mathbb{R}$, where $\operatorname{Re} T_i(\mathcal{N}_i) = \operatorname{Im} T_i(\mathcal{N}_i)$, by our convention.

The CVQFT-transformed channel transmission parameters are (upscaled by $\sqrt{n}$) expressed by the complex vector:

$$F(\mathbf{T}(\mathcal{N})) = \sum_{i=1}^{n} F(T_i(\mathcal{N}_i)) = \sum_{i=1}^{n}\sum_{k=1}^{n} T_k e^{\frac{-\mathrm{i}2\pi ik}{n}} \in \mathcal{C}^n, \qquad (48)$$

where $F(\mathbf{T}(\mathcal{N}))$ is the Fourier transform of (46). The $n$-dimensional $F(\Delta)$ complex vector is evaluated as

$$F(\Delta) = e^{\frac{-F(\Delta)^T \mathfrak{C}(F(\Delta))F(\Delta)}{2}} = e^{-\left[\frac{F(\Delta_1)^2 \sigma_{\mathcal{N}_1}^2 + \ldots + F(\Delta_n)^2 \sigma_{\mathcal{N}_n}^2}{2}\right]}, \qquad (49)$$

which is the Fourier transform of the $n$-dimensional zero-mean, circular symmetric complex Gaussian noise vector $\Delta \in \mathcal{CN}(0, \sigma_\Delta^2)_n$,

$$\Delta = (\Delta_1, \ldots, \Delta_n)^T \in \mathcal{CN}(0, \mathfrak{C}(\Delta)), \qquad (50)$$

where $\mathfrak{C}(\Delta) = \mathbb{E}[\Delta\Delta^\dagger]$, with independent, zero-mean Gaussian random components

$\Delta_{x_i} \in \mathbb{N}(0,\sigma^2_{\mathcal{N}_i})$, $\Delta_{p_i} \in \mathbb{N}(0,\sigma^2_{\mathcal{N}_i})$ with variance $\sigma^2_{\mathcal{N}_i}$, for each $\Delta_i$, which identifies the Gaussian noise of the $i$-th sub-channel $\mathcal{N}_i$ on the quadrature components in the phase space $\mathcal{S}$.
The CVQFT-transformed noise vector in (49) can be rewritten as

$$F(\Delta) = \left(F(\Delta_1),...,F(\Delta_n)\right)^T, \tag{51}$$

with independent components $F(\Delta_{x_i}) \in \mathbb{N}(0,\sigma^2_{F(\mathcal{N}_i)})$ and $F(\Delta_{p_i}) \in \mathbb{N}(0,\sigma^2_{F(\mathcal{N}_i)})$ on the quadratures, for each $F(\Delta_i)$. It also defines an $n$-dimensional zero-mean, circular symmetric complex Gaussian random vector $F(\Delta) \in \mathcal{CN}(0,\mathfrak{C}(F(\Delta)))$ with a covariance matrix

$$\mathfrak{C}(F(\Delta)) = \mathbb{E}\left[F(\Delta)F(\Delta)^\dagger\right], \tag{52}$$

and the noise variance $\sigma^2_{F(\mathcal{N})}$ of the independent Fourier-transformed quadratures is evaluated as

$$\begin{aligned}\sigma^2_{F(\mathcal{N})}\mathbf{I}_{n\times n} &= \tfrac{1}{n}\sum_{i=1}^{n}\sigma^2_{F(\mathcal{N}_i)} \\ &= \sigma^2_{F(\mathcal{N})} = \tfrac{1}{\sigma^2_\mathcal{N}},\end{aligned} \tag{53}$$

where $\sigma^2_\mathcal{N}$ is the noise variance of the Gaussian quantum channel $\mathcal{N}$, $\mathbf{I}_{n\times n}$ is the $n \times n$ identity matrix, hence $F(\Delta_i) \in \mathcal{CN}(0,\sigma^2_{F(\Delta_i)})$, and $\sigma^2_{F(\Delta_i)} = \mathbb{E}\left[|F(\Delta_i)|^2\right]$, with independent noise variance $\sigma^2_{F(\mathcal{N})}$ on the quadrature components (For simplicity, the notation of $\mathbf{I}_{n\times n}$ will be omitted from the description.). The LHS of (52) is justified by Proposition 1 and (33). It is an important corollary regarding the noise variance of the Fourier-transformed vector (51).
An AMQD *block* is formulated from $n$ Gaussian subcarrier continuous variables, as follows:

$$\mathbf{y}[j] = F(\mathbf{T}(\mathcal{N}))F(\mathbf{d})[j] + F(\Delta)[j],\ j=1,...,n, \tag{54}$$

where $j$ is the index of the AMQD block, $F(\mathbf{T}(\mathcal{N}))$ is defined in (48), $F(\mathbf{d}) = F(F^{-1}(\mathbf{z}))$, where $F^{-1}(\mathbf{z})$ is shown in (42), while

$$\begin{aligned}\mathbf{y}[j] &= (y_1[j],...,y_n[j])^T, \\ F(\mathbf{d})[j] &= (F(d_1)[j],...,F(d_n)[j])^T, \\ F(\Delta)[j] &= (F(\Delta_1)[j],...,F(\Delta_n)[j])^T.\end{aligned} \tag{55}$$

The squared magnitude

$$\tau = \|F(\mathbf{d})[j]\|^2 \tag{56}$$

identifies an exponentially distributed variable, with density $f(\tau) = \left(1/2\sigma^{2n}_\omega\right)e^{-\tau/2\sigma^2_\omega}$, and from the Parseval theorem [6] it follows that

$$\mathbb{E}[\tau] \leq n 2\sigma_\omega^2, \tag{57}$$

while the average quadrature modulation variance is

$$\sigma_\omega^2 = \tfrac{1}{n}\sum_{i=1}^{n} \sigma_{\omega_i}^2 = \sigma_{\omega_0}^2, \tag{58}$$

where $\sigma_{\omega_i}^2$ is the modulation variance of the quadratures of the subcarrier $|\phi_i\rangle$ transmitted by sub-channel $\mathcal{N}_i$.

The transformed vector **y** in (45) and (55) clearly demonstrates that the physical Gaussian channel is, in fact, divided into $n$ Gaussian quantum channels with independent noise variances. Each $\mathcal{N}_i$ Gaussian sub-channel is dedicated for the transmission of one Gaussian subcarrier CV from the $n$ subcarrier CVs.

∎

## 3.2 Gaussian Noise of the Sub-channels

Eve's optimal entangling cloner attack [2] in the multicarrier modulation setting is described as follows. Let the quadratures of the $i$-th subcarrier $|\phi_i\rangle$ transmitted by $\mathcal{N}_i$ be

$$(x_{in,i}, p_{in,i}),\ x_{in,i} \in \mathbb{N}(0, \sigma_{\omega_i}^2),\ p_{in,i} \in \mathbb{N}(0, \sigma_{\omega_i}^2), \tag{59}$$

where $\sigma_{\omega_i}^2$ is the modulation variance of the CVQFT transformed subcarrier CVs. Eve, equipped with $n$ EPR ancilla pairs $|\Psi_{EB}\rangle^{\otimes n}$ each with variance $W_i$, and prepares her $E_i$ as follows:

$$(x_{E,i}, p_{E,i}),\ x_{E,i} \in \mathbb{N}(0, \sigma_{\omega_i}^2 + \sigma_{\mathcal{N}_i}^2),\ p_{E,i} \in \mathbb{N}(0, \sigma_{\omega_i}^2 + \sigma_{\mathcal{N}_i}^2), \tag{60}$$

The part $B_i$ of is sent back to $\mathcal{N}_i$, which system has the following quadratures:

$$(x_{B,i}, p_{B,i}),\ x_{B,i} \in \mathbb{N}(0, \sigma_{\omega_i}^2 + \sigma_{\mathcal{N}_i}^2),\ p_{B,i} \in \mathbb{N}(0, \sigma_{\omega_i}^2 + \sigma_{\mathcal{N}_i}^2). \tag{61}$$

The simplified view of Eve's Gaussian attack in the multicarrier scenario is summarized in Fig. 2. Eve attacks each sub-channel with a BS with transmittance $T_{Eve,i} \in \mathcal{C}$, $0 < |T_{Eve,i}|^2 < 1$, and an entangled ancilla $|\Psi_{EB}\rangle$ with variance $W$. The quadratures of the $i$-th sub-channel are $(x_{in,i}, p_{in,i})$, Eve's quadratures are $(x_{E,i}, p_{E,i})$, Bob's received noisy quadratures are $(x'_{in,i}, p'_{in,i})$. Each sub-channel is characterized with a Gaussian noise $\mathbb{N}(0, \sigma_{\mathcal{N}_i}^2)$ on the quadrature components, with independent noise variance $\sigma_{\mathcal{N}_i}^2$. As shown in (50), Eve's optimal Gaussian attacks define an $n$-dimensional zero-mean, circular symmetric complex Gaussian random noise vector.

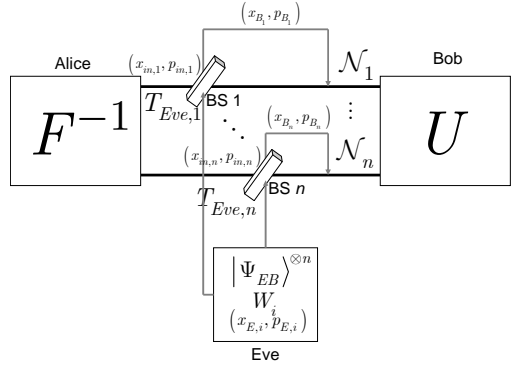

**Figure 2.** The entangler cloning attack in the multicarrier modulation scheme. The unitary refers to the CVQFT operation.

In the AMQD scenario, the appropriate noise vector is given by (49) and (52), as $F(\Delta) \in \mathcal{CN}(0, \mathfrak{C}(F(\Delta)))$, and $F(\Delta_i) \in \mathcal{CN}(0, \sigma^2_{F(\Delta_i)})$, respectively. For the security proof of AMQD against optimal Gaussian collective attacks is being presented in Theorem 4.

## 3.3 Crosstalk Noise on the Sub-channels

The crosstalk is an additive noise in multicarrier transmission that can occur between adjacent sub-channels. Crosstalk noise can result from a nonperfect sub-channel separation and sub-channel filtering procedure. As a result of interchannel crosstalk, some information from neighboring channels can leak, which acts as noise to the receiver. In practice, crosstalk arises from the imperfections of optical filters, optical switches or other optical components, e.g., from the imperfect isolation of different wavelength ports [19].

The effect of interchannel crosstalk in an AMQD modulation is depicted in Fig. 3. Crosstalk noise has no effect on the security of AMQD modulation, since it allows no more leaking of information to an eavesdropper than single-carrier CVQKD protocols (see Theorem 4).

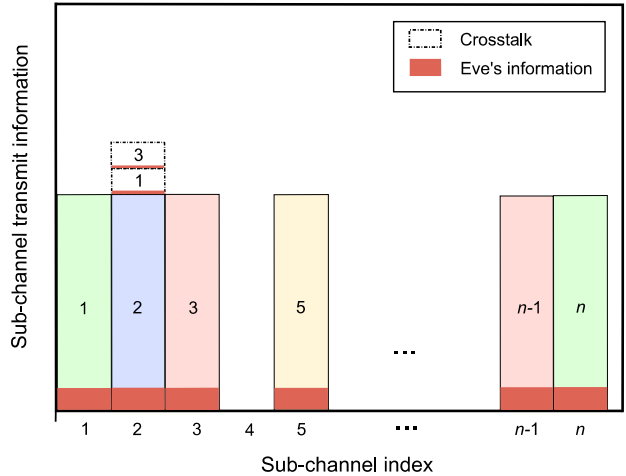

**Figure 3.** The effect of crosstalk noise between sub-channels. Due to the imperfections of practical devices, some information from neighboring sub-channels can be leaked to a given sub-channel, which acts as noise to the receiver.

As it is depicted in Fig. 3, for sub-channel 2, some crosstalk information brings up from neighboring sub-channels, 1 and 3. Any nonzero crosstalk information can leak information of only negligible valuable to Eve, as is proven in Theorem 4.

The crosstalk information $\gamma(\mathcal{N}_i)$ on the $i$-th sub-channel $\mathcal{N}_i$, $i = 1 \ldots n$, is evaluated as follows:

$$\gamma(\mathcal{N}_i) = \sum_{j \neq i} x_j \chi(A_j : B_j), \tag{62}$$

where $0 \leq x_j \leq 1$, $\chi(A_j : B_j) = H(B_j) - H(B_j | A_j)$ is the Holevo information of Alice and Bob conveyed by the neighboring sub-channel $\mathcal{N}_j$, for all $j \neq i$. The average crosstalk on the $n$ sub-channels is expressed as

$$\tilde{\gamma}(\mathcal{N}) = \tfrac{1}{n} \sum_n \sum_{j \neq i} x_j \chi(A_j : B_j). \tag{63}$$

Since the crosstalk noise $\gamma_i(\mathcal{N}_i)$ on sub-channel $\mathcal{N}_i$ acts for Bob as Gaussian noise (which is justified by the Central Limit Theorem, more precisely for $i \to \infty$), it can be modeled by $\gamma_i \in \mathcal{CN}(0, \sigma_{\gamma_i}^2)$ with variance $\sigma_{\gamma_i}^2 = \mathbb{E}[|\gamma_i|^2]$. This additional noise does not change the rate formulas of AMQD (see (74) and (80)), because this additional noise is already contained in the sub-channel's noise variance $\sigma_{\mathcal{N}_i}^2$.

In other words, the mutual information between Alice and Bob is completely characterized for a given $\mathcal{N}_i$ by the noise variance $\sigma_{\mathcal{N}_i}^2$, which can be decomposed as

$$\sigma_{\mathcal{N}_i}^2 = \sigma_{Eve,i}^2 + \sigma_{\gamma_i}^2, \tag{64}$$

where $\sigma_{Eve,i}^2$ is the noise variance of Eve's optimal Gaussian collective attack.

Assuming that $l$ sub-channels have been allocated for the multicarrier transmission, the mutual information function of Alice and Bob remains untouched under the presence of nonzero crosstalk, i.e., $I(A : B) = \tfrac{1}{l} \sum_l I(A_j : B_j)$.

However, the $\chi(B : E)$ Holevo information of Eve changes as follows:

$$\begin{aligned}\chi(B : E)_{\gamma_i} &= \chi(B : E) + \sum_n |T_{Eve,i}|^2 \gamma(\mathcal{N}_i) \\ &= \chi(B : E) + \sum_n \sum_{j \neq i} |T_{Eve,i}|^2 x_j \chi(A_j : B_j),\end{aligned} \tag{65}$$

where $|T_{Eve,i}|^2$ is the squared magnitude of Eve's $T_{Eve,i}$ normalized complex transmittance $T_{Eve,i} = \operatorname{Re} T_{Eve,i} + i \operatorname{Im} T_{Eve,i} \in \mathcal{C}$, where $0 \leq \operatorname{Re} T_{Eve,i} \leq 1/\sqrt{2}$, $0 \leq \operatorname{Im} T_{Eve,i} \leq 1/\sqrt{2}$ that characterizes the attack of sub-channel $\mathcal{N}_i$, and the quantity $\gamma(\mathcal{N}_i)$ is calculated with respect to the $n$ sub-channel (a nonzero crosstalk can also occur on an unused sub-channel, hence all sub-channels have to be taken into consideration in the RHS of (65)).

For detailed security proof of AMQD against optimal Gaussian collective attacks, see Theorem 4.

## 3.4 Run of Multicarrier Quadrature Division

The run of the multicarrier quadrature division is sketched as follows. In the initial phase, Alice draws an $n$-dimensional, zero-mean circular symmetric complex Gaussian random vector $\mathbf{z} = \mathbf{x} + i\mathbf{p} = (z_1,...,z_n)^T \in \mathcal{CN}(0, \mathbf{K_z})$, $z_i = p_i + iq_i$, and $x_i \in \mathbb{N}(0, \sigma^2_{\omega_0})$, $p_i \in \mathbb{N}(0, \sigma^2_{\omega_0})$ are i.i.d. Gaussian random variables that identifies the $x$ position and $p$ momentum quadratures in the phase space $\mathcal{S}$, while $\sigma^2_{\omega_0}$ is the modulation variance (at a single-carrier transmission).

In the next step, Alice applies the inverse FFT on $\mathbf{z}$, that gives her the results of an $n$-dimensional, zero-mean circular symmetric complex Gaussian random vector $\mathbf{d} = \mathbf{x} + i\mathbf{p} = (d_1,...,d_n)^T \in \mathcal{CN}(0, \mathbf{K_d})$. According to $\mathbf{d}$, she prepares the $|\phi_{1...n}\rangle$ Gaussian sub-carrier CVs, by modulating with $\sigma^2_\omega \neq \sigma^2_{\omega_0}$ the position and momentum quadratures, where $|\phi_i\rangle$ is the $i$-th subcarrier continuous variable. The $n$ subcarrier coherent states $|\phi_i\rangle$ divide the physical Gaussian quantum channel into $n$ physical Gaussian quantum channels, each equipped with an independent noise variance $\sigma^2_{\mathcal{N}_i}$.

In the decoding phase, Bob applies the CVQFT unitary operation $U$ on the received noisy Gaussian subcarrier CVs, $|\phi'_i\rangle$, which results him the noisy coherent state versions of Alice's Gaussian variables, $|\varphi'_{1...n}\rangle = |z'_{1...n}\rangle = |\mathbf{z}'\rangle$, and the Fourier transformed sub-channel noise variance $\sigma^2_{F(\mathcal{N}_i)}$. The CVQFT-transformed $|F(T_i(\mathcal{N}_i))|^2$ transmission parameters of the Gaussian sub-channels are strongly diverse (this will be shown in Section 4), which makes available the use of an *adaptive variance modulation* to improve the tolerable noise and excess noise.

The steps of multicarrier quadrature division modulation are summarized in Fig. 4.

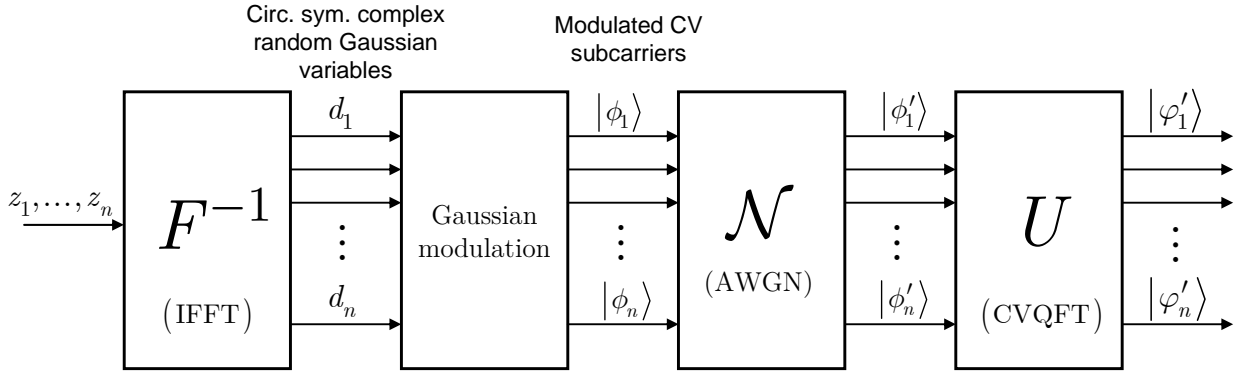

**Figure 4.** The AMQD modulation scheme. Alice draws an $n$-dimensional, zero-mean, circular symmetric complex Gaussian random vector $\mathbf{z}$, which are then inverse Fourier-transformed by $F^{-1}$. The resulting vector $\mathbf{d}$ encodes the subcarrier quadratures for the Gaussian modulation. In the decoding, Bob applies the $U$ unitary CVQFT on the $n$ subcarriers to recover the noisy version of Alice's original variable as a continuous variable in the phase space.

(*Note*: In the two-way protocol, Bob sends the subcarrier CVs to Alice, who generates coupled

Gaussian CVs with her BS. The resulting Gaussian subcarrier CV is then sent back to Bob, who applies the CVQFT operation. Alice also applies a CVQFT operation on her system.) The reason behind the improvement is that the dB limits of the transmission over the Gaussian quantum channel can significantly be extended by the Fourier-transformed multicarrier continuous variables. The adaptive modulation sends information through only the high-quality sub-channels, while the noisy sub-channels transmit no information, as it is proposed in the next section.

## 4 Adaptive Modulation Variance

**Theorem 2.** (Adaptive modulation variance.) *The ratio* $\nu_i = \sigma_{\mathcal{N}}^2 / |F(T_i(\mathcal{N}_i))|^2$, *where* $|F(T_i(\mathcal{N}_i))|^2 = \left|\sum_{k=1}^{n} T_k e^{\frac{-i2\pi ik}{n}}\right|^2$, $i = 1...n$, *provides the* $\sigma_\omega^2 = \nu_{Eve} - \min(\nu_i)$ *optimal constant modulation variance for the Gaussian sub-channels, where* $\nu_{Eve}$ *is a security bound of the optimal Gaussian collective attack.*

*Proof.*
The proof consists of two parts. First, we show that there exists an optimal constant modulation variance for the sub-channels with average $\sigma_\omega^2 = \frac{1}{n}\sum_n \sigma_{\omega_i}^2 = \nu_{Eve} - (\nu)$, where $\nu_{Eve}$ is the security bound of Eve's optimal Gaussian collective attack, and $\nu$ is

$$\nu = \frac{1}{n}\sum_n \nu_i = \frac{1}{n} \frac{\sum_{i=1}^{n} \sigma_{\mathcal{N}_i}^2}{\sum_{i=1}^{n}\left|\sum_{k=1}^{n} T_k e^{\frac{-i2\pi ik}{n}}\right|^2} = \frac{\sigma_{\mathcal{N}}^2}{\sum_{i=1}^{n}|F(T_i(\mathcal{N}_i))|^2}. \tag{66}$$

Then, we show that for low SNRs, the optimal solution is $\nu = \min(\nu_i)$, which requires no the exact knowledge of the state of the sub-channels.

The modulation variances of each of the $\mathcal{N}_i$ sub-channels (zero or nonzero) are dependent on the value of $\nu_{Eve}$. This parameter is defined as

$$\nu_{Eve} = \tfrac{1}{\lambda}, \tag{67}$$

where $\lambda$ is the *Lagrange multiplier* as

$$\lambda = \left|F(T_{\mathcal{N}}^*)\right|^2 = \tfrac{1}{n}\sum_{i=1}^{n}\left|\sum_{k=1}^{n} T_k^* e^{\frac{-i2\pi ik}{n}}\right|^2, \tag{68}$$

where $T_{\mathcal{N}}^*$ is the *expected* transmittance of the *n* sub-channels under an optimal Gaussian attack. From $\lambda$, and the $\sigma_{\omega_i}^2$ modulation variances of the $\mathcal{N}_i$ sub-channels, a Lagrangian can be constructed as

$$\mathcal{L}\left(\lambda, \sigma_{\omega_1}^2 ... \sigma_{\omega_n}^2\right) = \sum_{i=1}^{n} \log_2\left(1 + \frac{\sigma_{\omega_i}^2 |F(T_i(\mathcal{N}_i))|^2}{\sigma_{\mathcal{N}}^2}\right) - \lambda\sum_{i=1}^{n} \sigma_{\omega_i}^2. \tag{69}$$

By the Kuhn-Tucker condition [6], [9-10], follows that $\frac{\partial \mathcal{L}}{\partial \sigma_{\omega_i}^2} = 0$ if only the $i$-th sub-channel gets a non-zero modulation variance, $\sigma_{\omega_i}^2 > 0$, while $\frac{\partial \mathcal{L}}{\partial \sigma_{\omega_i}^2} \leq 0$, if the sub-channel gets zero modulation variance, $\sigma_{\omega_i}^2 = 0$ [6], [9-10].

After some calculations, one gets the following average modulation variance:

$$\sigma_\omega^2 = \frac{1}{n}\sum_{i=1}^{n}\left(\nu_{Eve} - \frac{\sigma_\mathcal{N}^2}{\left|F(T_i(\mathcal{N}_i))\right|^2}\right) = \frac{1}{n}\sum_{i=1}^{n}\left(\nu_{Eve} - \nu_i\right), \qquad (70)$$

where $\nu$ is shown in (66). One can readily see that in (70), each sub-channel is allocated by a different modulation variance, depending on the actual value of $\left|F(T_i(\mathcal{N}_i))\right|^2$. The reason for this is as follows. Only those $l < n$ sub-channels can transmit information for which $\nu_i < \nu_{Eve}$; otherwise, the channel gets zero modulation variance. (In general, this kind of strategy is called water-filling [6], [9-10].) Since it is not a reasonable assumption in a practical CVQKD that the transmitter would have an *exact* knowledge about the state of each Gaussian sub-channels, at this point we have to introduce a more flexible technique. Our answer is as follows.

In fact, it is not a required condition to calculate with the exact $\nu_i$ parameters and modulation variances $\sigma_{\omega_i}^2$ for the sub-channels. A simplified solution exists: give a *constant* modulation variance $\sigma_\omega^2$ for those $l$ $\mathcal{N}_i$ sub-channels, for which $\nu_i < \nu_{Eve}$ is satisfied:

$$\sigma_\omega^2 = \frac{1}{l}\sum_{i=1}^{l}\left(\nu_{Eve} - \frac{\sigma_\mathcal{N}^2}{\max_i\left|F(T_i(\mathcal{N}_i))\right|^2}\right) = \nu_{Eve} - \min(\nu_i). \qquad (71)$$

It is particular convenient, since at a given $\nu_{Eve}$ bound, it is enough to find a given $\max_i\left|F(T_{1...l}(\mathcal{N}_{1...l}))\right|^2$ for those $\mathcal{N}_i$ sub-channels, for which $\nu_i < \nu_{Eve}$ hold (*Note*: It can be determined in a pre-calibration phase prior to the main run of the protocol. In practice, this phase can be established in one simple step by sending empty "pilot" continuous variables over the sub-channels that contain no valuable information.), and then allocating the same modulation variance $\sigma_\omega^2$ for all of these sub-channels.

Particularly, the significance of the adaptive-variance modulation proposed in (71) is crucial for low SNRs, which is precisely the case in a long-distance scenario, since the information transmission capability of the Gaussian sub-channels become very sensitive in the low SNR regimes [4], [6], [9-10]. At low SNRs the constant allocation provides an optimal solution [9-10], because its performance is very close to the exact allocation and can be performed with no exact knowledge about the state of the sub-channels. This is very good news, because the proposed AMQD modulation scheme allocates a constant modulation variance for the good Gaussian sub-channels.

The algorithm of the optimal *constant* [9-10] modulation variance adaption is summarized as follows.

**Algorithm**

1. Let the squared magnitudes of the Fourier-transformed sub-channel transmittance coefficients given in an ordered list $L = |F(T_{1...l}(\mathcal{N}_{1...l}))|^2$ so that $|F(T_i(\mathcal{N}_i))|^2 \geq |F(T_{i+1}(\mathcal{N}_{i+1}))|^2$ and $\nu_i \leq \nu_{i+1}$, and let $\lambda$ be the Lagrange coefficient and $\nu_{Eve} = 1/\lambda$. Let $\chi$ be the largest index in the list, denoted by $L(\chi)$, for which $|F(T_\chi(\mathcal{N}_\chi))|^2 > \lambda$. For $L(i > \chi)$, $|F(T_{\chi+1...n}(\mathcal{N}_{\chi+1...n}))|^2 \leq \lambda$.

2. Let $\sigma_\omega^2$ be the average modulation $\sigma_\omega^2 = \max_{\sigma_{\omega_i}^2} \sum_{i=1}^{\chi} \log_2\left(1 + \frac{\sigma_{\omega_i}^2 |F(T_i(\mathcal{N}_i))|^2}{\sigma_{\mathcal{N}_i}^2}\right)$, where $\sigma_{\omega_i}^2 \geq 0$. Determine the constant modulation variance $\sigma_\omega^2$ as $\sigma_\omega^2 = \frac{1}{\chi}\sum_\chi \sigma_{\omega_i}^2$.

3. If $\nu_{\chi+1} \geq \sigma_\omega^2 + \nu_1$ then $L(\chi) = L(\chi - 1)$, and re-determine $\sigma_\omega^2$ as $\sigma_\omega^2 = \frac{1}{\chi-1}\sum_{\chi-1}\sigma_{\omega_i}^2$.

4. If $\nu_{\chi+1} < \sigma_\omega^2 + \nu_1$, then $L(\chi) = L(\chi+1)$, compute $\sum_{i=1}^{\chi+1}\log_2\left(1 + \frac{\sigma_\omega^2 |F(T_i(\mathcal{N}_i))|^2}{\sigma_\mathcal{N}^2}\right)$, where $|F(T_i(\mathcal{N}_i))|^2 = \left|\sum_{k=1}^n T_k e^{\frac{-i2\pi ik}{n}}\right|^2$.

The rate over the $\chi+1$ Gaussian sub-channels is $R = \sum_{\chi+1}\log_2\left(1 + \frac{\sigma_\omega^2 |F(T_i(\mathcal{N}_i))|^2}{\sigma_\mathcal{N}^2}\right)$, which approximates the complex domain capacity by precisely $\sum_{\chi+1} 2^{-\log_2\left(1+\frac{\sigma_\omega^2 |F(T_i(\mathcal{N}_i))|^2}{\sigma_\mathcal{N}^2}\right)}\Big/\ln 2$ [9-10].

The modulation variance adaption scheme is summarized in Fig. 5. The parameter $\nu_i$ of the Gaussian sub-channels is depicted in yellow. If $\nu_i$ is under a critical limit $\nu_{Eve}$, the channel is assumed to be useful and can be used for information transmission. If $\nu_i < \nu_{Eve}$, Alice allocates a constant modulation variance $\sigma_{\omega_i}^2 = \sigma_\omega^2$ according to (71), for the Gaussian sub-channel $\mathcal{N}_i$. If $\nu_i \geq \nu_{Eve}$, Alice allocates zero modulation variance for $\mathcal{N}_i$, i.e., $\sigma_{\omega_i}^2 = 0$.

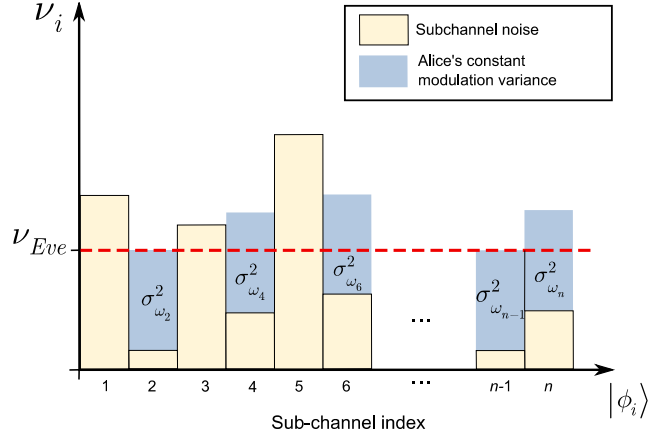

**Figure 5.** The constant modulation variance allocation mechanism. If the $i$-th sub-channel $\mathcal{N}_i$ is very noisy, i.e., $\nu_i \geq \nu_{Eve}$, Alice will not use that sub-channel, i.e., the modulation variance $\sigma^2_{\omega_i}$ of $|\phi_i\rangle$ is 0. Only those sub-channels will be used for the transmission for which $\nu_i$ is under the critical bound $\nu_{Eve}$ (red dashed line). Assuming $l$ sub-channels with $\nu_i < \nu_{Eve}$, the modulation variance for these sub-channels is chosen to be a constant $\sigma^2_{\omega_i} = \nu_{Eve} - \min(\nu_1,\ldots,\nu_n)$, where $\sum_l \sigma^2_{\omega_i} = l\sigma^2_\omega < n\sigma^2_{\omega_0}$, and $\sigma^2_{\omega_0}$ is the single-carrier modulation variance.

The AMQD is equipped with all of those properties that allow it to meet the requirements of an experimental protocol, since its real potential is brought to life at very low SNRs. As a fine corollary, the transmission efficiency significantly can be boosted in an experimental long-distance CVQKD scenario.

The adaptive modulation variance can be rephrased in terms of the normalized $\frac{1}{n}|T(\mathcal{N})|^2$ squared magnitude of the channel transmittance $T(\mathcal{N})$, where $n$ is the number of subcarrier CVs. Let $\nu = \sigma^2_\mathcal{N} \Big/ \Big|F\Big(\frac{1}{n}|T(\mathcal{N})|^2\Big)\Big|^2$, and $\sigma^2_\omega\Big(\frac{1}{n}|T(\mathcal{N})|^2\Big) = \nu_{Eve} - \frac{\sigma^2_\mathcal{N}}{\max\Big|F\Big(\frac{1}{n}|T(\mathcal{N})|^2\Big)\Big|^2}$. In this approach, the optimal modulation variance is

$$\int_0^{\frac{1}{2}|T(\mathcal{N})|^2} \sigma^2_\omega(x)\,dx = \sigma^2_\omega, \tag{72}$$

where $x = \frac{1}{n}|T(\mathcal{N})|^2$, which precisely coincidences with (71).

In Fig. 6, the adaptive modulation variance technique in terms of the normalized quantity [6] $\frac{1}{n}|T(\mathcal{N})|^2$ and parameter $\nu = \sigma^2_\mathcal{N} \Big/ \Big|F\Big(\frac{1}{n}|T(\mathcal{N})|^2\Big)\Big|^2$ is shown. The maximal value of $|F(T)|^2$ is obtained at $\frac{1}{n}|T(\mathcal{N})|^2 \to 0$, because as $n$ increases, the normalized quantity represents a finer sampling of $T(\mathcal{N})$ of the physical quantum channel.

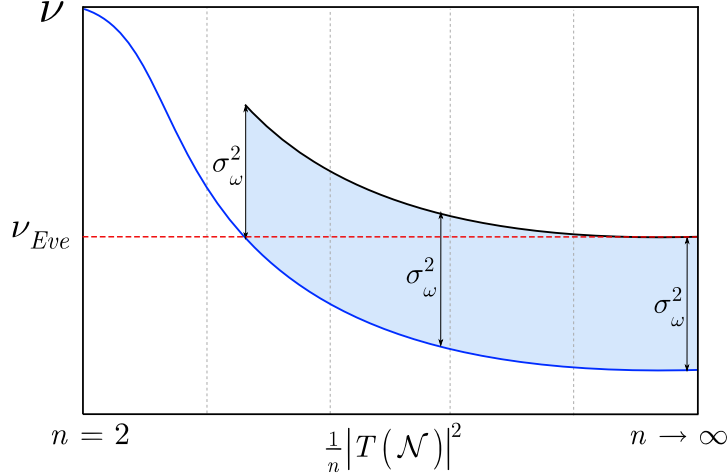

**Figure 6.** The constant modulation variance allocation in function of the normalized squared magnitude. As the number of subcarrier goes to infinity, the rate goes arbitrary close to the real capacity of the Gaussian quantum channel. Only those sub-channels get nonzero modulation variance $\sigma_\omega^2$ for which $\nu_i < \nu_{Eve}$ holds.

For the $l$ Gaussian sub-channels $\mathcal{N}_i$, $\nu_i < \nu_{Eve}$, the mutual information is evaluated as follows. Assuming a single-carrier modulation over a Gaussian channel $\mathcal{N}$ with noise variance $\sigma_\mathcal{N}^2$, and modulation variance $\sigma_{\omega_0}^2$, let the $R_{single}(\mathcal{N})$ maximal rate between Alice and Bob is

$$R_{single}(\mathcal{N}) = \log_2\left(1 + \frac{\sigma_{\omega_0}^2 |T(\mathcal{N})|^2}{\sigma_\mathcal{N}^2}\right). \tag{73}$$

For the $l$ Gaussian sub-channels $R_{AMQD}(\mathcal{N})$ is evaluated as

$$R_{AMQD}(\mathcal{N}) = \max_{\forall i} \sum_{i=1}^{l} \log_2\left(1 + \frac{\sigma_{\omega_i}^2 |F(T_i(\mathcal{N}_i))|^2}{\sigma_\mathcal{N}^2}\right), \tag{74}$$

where $\sigma_\mathcal{N}^2 = \sigma_\mathcal{N}^2 \mathbf{I}_{l \times l} = \frac{1}{l}\sum_{i=1}^{l} \sigma_{\mathcal{N}_i}^2$ follows from (52) for the noise variance of the $l$ Gaussian sub-channels. For the $\sigma_\omega^2$ averaged modulation variance of the $l$ sub-channels, the relation

$$\sigma_\omega^2 = \frac{1}{l}\sum_{i=1}^{l} \sigma_{\omega_i}^2 < \sigma_{\omega_0}^2 \tag{75}$$

follows, see (58). The multicarrier modulation transmits the same amount of information, hence at (75), $\sigma_{\omega_0}^2 |T(\mathcal{N})|^2 = \frac{1}{l}\sum_l \sigma_{\omega_i}^2 |F(T(\mathcal{N}_i))|^2$ holds, i.e.,

$$R_{AMQD}(\mathcal{N}) = R_{single}(\mathcal{N}). \tag{76}$$

Furthermore, (76) can be increased by improving the average modulation variance in (75) up to $\sigma_\omega^2 = \sigma_{\omega_0}^2$. In this case, between (73) at $\sigma_{\omega_0}^2$, and (74) at $\sigma_\omega^2 = \sigma_{\omega_0}^2$, the connection is $\sigma_{\omega_0}^2 |T(\mathcal{N})|^2 < \frac{1}{l}\sum_l \sigma_{\omega_i}^2 |F(T(\mathcal{N}_i))|^2$, hence:

$$R_{AMQD}(\mathcal{N}) > R_{single}(\mathcal{N}), \tag{77}$$

which shows that better rates can be reached by the AMQD modulation at a given noise variance $\sigma_\mathcal{N}^2$ and modulation variance $\sigma_{\omega_0}^2$, in comparison to the single-carrier modulation.

Let $0 \leq \Omega \leq 1$ be the probability that the $\sum_l |F(T_i(\mathcal{N}_i))|^2$ sum of the squared magnitude of the Fourier transformed channel transmission coefficients of the $l$ Gaussian sub-channels pick up the maximum $\sum_l |F(T_i(\mathcal{N}_i))|^2 = \max_{\forall i} \sum_l |F(T_i(\mathcal{N}_i))|^2$:

$$\Omega = \Pr\left(\sum_l |F(T_i(\mathcal{N}_i))|^2 = \max_{\forall i} \sum_l |F(T_i(\mathcal{N}_i))|^2\right). \tag{78}$$

From this, the achievable rate is

$$R_{AMQD}(\mathcal{N}) = \Omega \log_2\left(1 + \frac{\max_i \sum_l |F(T_i(\mathcal{N}_i))|^2}{\Omega} \cdot \text{SNR}\right). \tag{79}$$

It can be concluded that the adaptive modulation variance increases the performance, especially in low SNR regimes, in which situation a constant modulation variance for all sub-channels is optimal. It also optimizes the received modulation variance at the decoder.

Finally, in terms of (72) the rate is evaluated as

$$R_{AMQD}(\mathcal{N}) = \int_0^{\frac{1}{2}|T(\mathcal{N})|^2} \log_2\left(1 + \frac{\sigma_\omega^2(x)\left|F\left(\frac{1}{n}|T(\mathcal{N})|^2\right)\right|^2}{\sigma_\mathcal{N}^2}\right) dx, \tag{80}$$

where $x = \frac{1}{n}|T(\mathcal{N})|^2$ and $n \to \infty$.

These results conclude the proof of Theorem 2. ∎

## 5 Efficiency of AMQD Modulation

**Theorem 3** (Efficiency of the AMQD modulation). *At a given channel transmittance $|T(\mathcal{N})|^2$, more loss can be tolerated over the Gaussian sub-channels, which leads to higher tolerable excess noise at a given modulation variance $\sigma_{\omega_0}^2$.*

*Proof.*

The efficiency of the adaptive multicarrier division technique can be approached by how efficiently the allocated modulation frequencies are utilized at a given secret key rate. Assuming a single-carrier quadrature encoding-scheme with efficiency

$$\eta_{single} = \frac{bR}{\sigma_{\omega_0}^2}, \qquad (81)$$

where $b = 2$ is the number of quadrature bases (position and momentum bases), $R$ is the secret key rate, and $\sigma_{\omega_0}^2$ is the modulation variance used for the modulation of $n$ coherent states. In the adaptive multicarrier division technique, the rate $R$ given in (81) can be reached with efficiency:

$$\eta_{AMQD} = \frac{bR}{\sigma_{\omega}^2}, \qquad (82)$$

where $\sigma_{\omega}^2$ is the constant modulation variance is used for the $l$ subcarriers. If $\sigma_{\omega}^2 < \sigma_{\omega_0}^2$ holds at a given $R$, the multicarrier modulation scheme transmits the same amount of information such as the single-carrier modulation scheme, with increased efficiency, $\eta_{AMQD} > \eta_{single}$.

The improvement in the tolerable loss is as follows. In the single-carrier modulation scheme, at the 3.1 dB of signal attenuation the quantum channel efficiency falls below 0.5. In the multicarrier modulation, the tolerable loss is more than 3.1 dB at the same modulation variance $\sigma_{\omega_0}^2$ (justified by the convention of full width at half maximum (FWHM)).

For $\sigma_{\omega_0}^2$ and AMQD modulation, the -3.1 dB attenuation brings up at diverse amount of loss, which is equivalent to the use of an improved *virtual* modulation variance $A\sigma_{\omega_0}^2$, where $A > 1$. At $\sigma_{\omega_0}^2$, a higher modulation variance $A\sigma_{\omega_0}^2$ can be virtually simulated by the multicarrier transmission, which result higher tolerable loss (dBs) and higher tolerable excess noise in overall. To demonstrate it, let the input Gaussian

$$g_1(x) = \frac{1}{\sigma_{\omega_0}\sqrt{2\pi}} e^{\frac{-x^2}{2\sigma_{\omega_0}^2}}. \qquad (83)$$

The $\sigma_{\omega}^2 < \sigma_{\omega_0}^2$ modulation variance of the $l$ sub-channels leads to a Gaussian

$$g_2(x) = \frac{1}{\sigma_{\omega}\sqrt{2\pi}} e^{\frac{-x^2}{2\sigma_{\omega}^2}}. \qquad (84)$$

From (84) the *virtual* Gaussian signal at a modulation variance $\sigma_{\omega_0}^2$ is as follows

$$g_3(x) = \frac{1}{\sqrt{A}\sigma_{\omega_0}\sqrt{2\pi}} e^{\frac{-x^2}{2A\sigma_{\omega_0}^2}}, \qquad (85)$$

where $A\sigma_{\omega_0}^2 > \sigma_{\omega_0}^2$, and $A > 1$, which is evaluated as

$$A = \frac{\frac{1}{l}\sum_l |F(T_i(\mathcal{N}_i))|^2}{|T(\mathcal{N})|^2} = \frac{\frac{1}{l}\sum_l |F(T_i(\mathcal{N}_i))|^2}{2T(\mathcal{N})^2} = \frac{\frac{1}{l}\sum_{i=1}^{l}\left|\sum_{k=1}^{l} T_k e^{\frac{-i2\pi ik}{l}}\right|^2}{2T(\mathcal{N})^2}. \tag{86}$$

The FWHMs of these Gaussian signals are as follows:

$$\mathrm{FWHM}\big(g_1(x)\big) = 2\sqrt{2\ln 2}\sigma_{\omega_0}, \tag{87}$$

$$\mathrm{FWHM}\big(g_2(x)\big) = 2\sqrt{2\ln 2}\sigma_{\omega} \tag{88}$$

and

$$\mathrm{FWHM}\big(g_3(x)\big) = \sqrt{A}\,2\sqrt{2\ln 2}\sigma_{\omega_0}. \tag{89}$$

These results are summarized in Fig. 7.

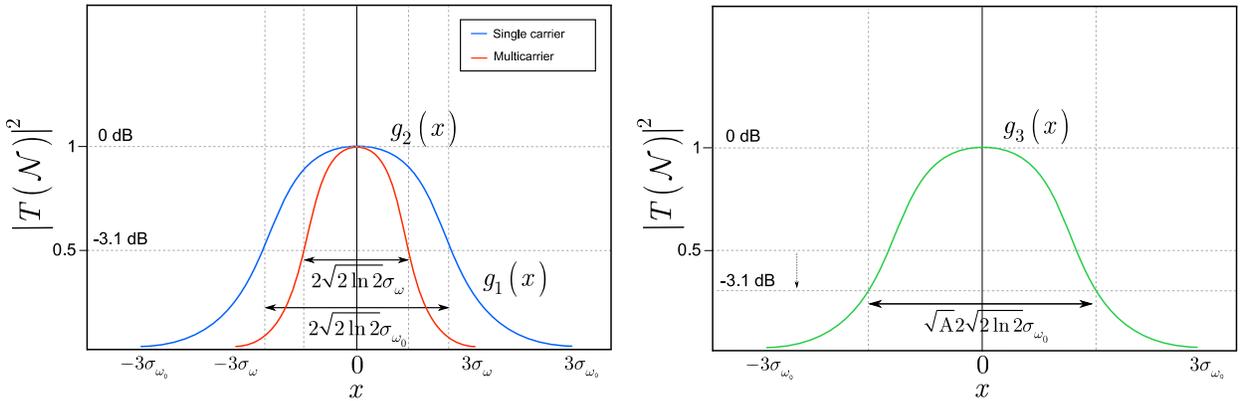

**Figure 7.** (a). The FWHM of the single-carrier Gaussian signal (blue) and the Gaussian signal of the subcarrier transmission (red). (b). The FWHM of the virtual Gaussian signal at a modulation variance of $\sigma_{\omega_0}^2$ is $\sqrt{A}\,2\sqrt{2\ln 2}\sigma_{\omega_0}$, $A > 1$. The AMQD allows more tolerable loss at a given modulation variance $\sigma_{\omega_0}^2$.

The improvement in the *excess noise* is as follows [3]. In the single-carrier case, at a given $T_{Eve}$ and $W$, the *excess noise* is as follows [3]:

$$N_{single} = \frac{(W-1)\big(|T_{Eve}|^2\big)}{1-|T_{Eve}|^2}, \tag{90}$$

where $|T_{Eve}|^2 < |T|^2 \in [0,1]$ is the squared magnitude of the complex variable $T_{Eve}$.
In the AMQD modulation, at a given $W$, the excess noise is reduced to

$$N_{AMQD} = \frac{(W-1)\big(\frac{1}{n}\sum_n |F(T_{Eve,i})|^2\big)}{1-\frac{1}{n}\sum_n |F(T_{Eve,i})|^2}, \tag{91}$$

where

$$F\left(T_{Eve,i}\right) = \sum_{k=1}^{n} T_{Eve,k} e^{\frac{-i2\pi ik}{n}} \tag{92}$$

and

$$\tfrac{1}{n}\sum_n \left|F\left(T_{Eve,i}\right)\right|^2 = \tfrac{1}{n}\sum_n \left|\sum_{k=1}^{n} T_{Eve,k} e^{\frac{-i2\pi ik}{n}}\right|^2 \leq \left|T_{Eve}\right|^2 < \left|T\right|^2 \leq \tfrac{1}{n}\sum_n \left|F\left(T_i\right)\right|^2, \tag{93}$$

which relation is justified by the CVQFT transformation. Since, at a given $W$, the tolerable excess noise depends only on $T_{Eve}$, from (93) it follows, that in the AMQD modulation scheme higher amount of excess noise $N$ can be tolerated.

The improvement in the tolerable excess noise can be approached by the ratio $\kappa \geq 1$ between the excess noise $N_{single}$ and $N_{AMQD}$ as:

$$\begin{aligned}
\kappa &= \frac{N_{single}}{N_{AMQD}} \\
&= \frac{(W-1)\left(\left|T_{Eve}\right|^2\right)}{1-\left|T_{Eve}\right|^2} \frac{1-\tfrac{1}{n}\sum_n \left|F\left(T_{Eve,i}\right)\right|^2}{(W-1)\left(\tfrac{1}{n}\sum_n \left|F\left(T_{Eve,i}\right)\right|^2\right)} \\
&= \frac{\left|T_{Eve}\right|^2 - \left|T_{Eve}\right|^2 \tfrac{1}{n}\sum_n \left|F\left(T_{Eve,i}\right)\right|^2}{\tfrac{1}{n}\sum_n \left|F\left(T_{Eve,i}\right)\right|^2 - \left|T_{Eve}\right|^2 \tfrac{1}{n}\sum_n \left|F\left(T_{Eve,i}\right)\right|^2} \\
&= \frac{\left|T_{Eve}\right|^2 - \left|T_{Eve}\right|^2 \tfrac{1}{n}\sum_n \left|F\left(T_{Eve,i}\right)\right|^2}{\left|T\right|^2 \tfrac{1}{n}\sum_n \left|F\left(T_{Eve,i}\right)\right|^2}.
\end{aligned} \tag{94}$$

The $N_{tol}$ tolerable excess noise in AMQD modulation with homodyne measurement [3] is depicted in Fig. 8. The improvement in the tolerable excess noise is $N_{tol,AMQD} = \alpha N_{tol,single}$, where $\alpha = x\kappa = x\, N_{single}/N_{AMQD} \geq 1$, and $N_{AMQD}/N_{single} \leq x \leq 1$.

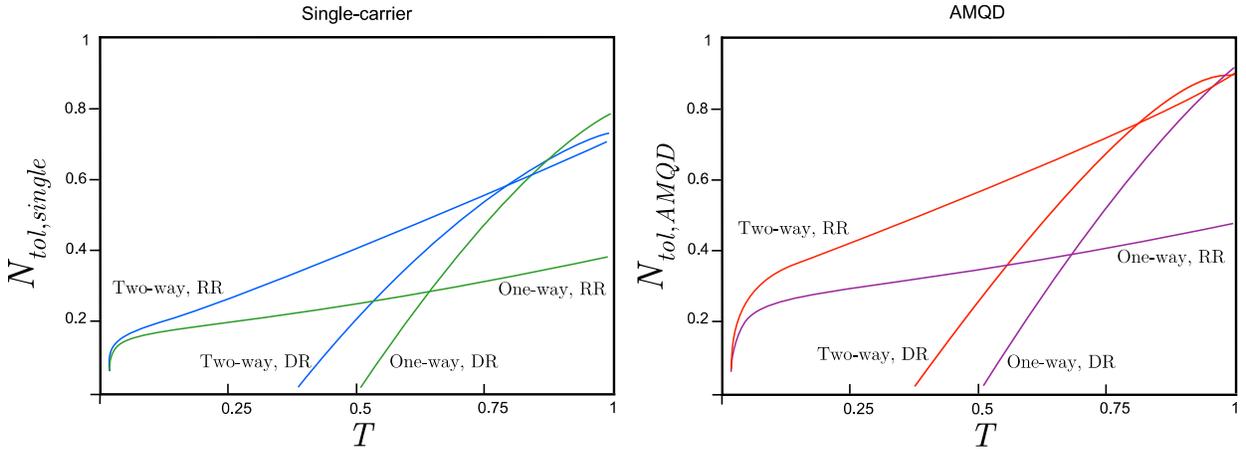

**Figure 8.** Tolerable excess noise in function of channel transmittance in a single-carrier Gaussian modulation with homodyne measurement (a) and in AMQD modulation (b). Abbreviations: DR – Direct Reconciliation, RR – Reverse Reconciliation.

The improvement in the amount of tolerable excess noise is more significant in two-way CVQKD, since the multicarrier transmission is present in both directions between Alice to Bob.

∎

The improvements of the AMQD modulation lead to different secret key rates, as it is summarized in Section 5.1.

## 5.1 Secret Key Rates in an AMQD Modulation

The secret key rates of AMQD modulation are evaluated as follows. Let $n$ subcarriers, and $l$ Gaussian sub-channels with $\nu_i < \nu_{Eve}$, and let the variance of Eve's EPR-ancilla $W$ [2-3].

The CVQKD secret key rates with AMQD modulation follows from the rates already obtained in [3], and sketched as follows (The rates are measured in the real dimension which is the reason behind the normalization of $1/2$.). The detailed proof of the following rate formulas can be found in [3]; thus, for simplicity, we omit these steps here.

### 5.1.1 AMQD Modulation in One-way CVQKD

*5.1.1.1 Homodyne measurement, Reverse Reconciliation*

$$R_{one-way}^{RR,\mathrm{hom}} = \tfrac{1}{2}\log_2 \frac{\left(\tfrac{1}{l}\sum_l |F(T_i(\mathcal{N}_i))|^2\right)e}{\left(1-\left(\tfrac{1}{l}\sum_l |F(T_i(\mathcal{N}_i))|^2\right)\right)b} \\ - \left(\frac{\sqrt{Wb/e}+1}{2}\log_2 \frac{\sqrt{Wb/e}+1}{2} - \frac{\sqrt{Wb/e}-1}{2}\log_2 \frac{\sqrt{Wb/e}-1}{2}\right) \\ - \left(\frac{W+1}{2}\log_2 \frac{W+1}{2} - \frac{W-1}{2}\log_2 \frac{W-1}{2}\right), \tag{95}$$

where

$$b = \left(1 - \left(\tfrac{1}{l}\sum_l |F(T_i(\mathcal{N}_i))|^2\right)\right)W + \tfrac{1}{l}\sum_l |F(T_i(\mathcal{N}_i))|^2, \tag{96}$$

$$e = \left(1 - \tfrac{1}{l}\sum_l |F(T_i(\mathcal{N}_i))|^2\right) + \left(\tfrac{1}{l}\sum_l |F(T_i(\mathcal{N}_i))|^2\right)W. \tag{97}$$

*5.1.1.2 Homodyne measurement, Direct Reconciliation*

$$R_{one-way}^{DR,\mathrm{hom}} = \tfrac{1}{2}\log_2 \frac{W}{\left(1-\tfrac{1}{l}\sum_l |F(T_i(\mathcal{N}_i))|^2\right)b} - \left(\frac{W+1}{2}\log_2 \frac{W+1}{2} - \frac{W-1}{2}\log_2 \frac{W-1}{2}\right). \tag{98}$$

### 5.1.2 AMQD Modulation in Two-way CVQKD

*5.1.2.1 Homodyne measurement, Reverse Reconciliation*

$$R_{two-way}^{RR,\mathrm{hom}} = \tfrac{1}{2}\log_2 \frac{\left[1-\left(\tfrac{1}{l}\sum_l |F(T_i(\mathcal{N}_i))|^2\right)+\left(\tfrac{1}{l}\sum_l |F(T_i(\mathcal{N}_i))|^2\right)^2\right]}{\left(1-\tfrac{1}{l}\sum_l |F(T_i(\mathcal{N}_i))|^2\right)^2} - \left(\frac{W+1}{2}\log_2 \frac{W+1}{2} - \frac{W-1}{2}\log_2 \frac{W-1}{2}\right). \tag{99}$$

*5.1.2.2 Homodyne measurement, Direct Reconciliation*

$$R_{two-way}^{DR,\mathrm{hom}} = \tfrac{1}{2}\log_2 \frac{\tfrac{1}{l}\sum_l |F(T_i(\mathcal{N}_i))|^2}{\left(1-\tfrac{1}{l}\sum_l |F(T_i(\mathcal{N}_i))|^2\right)^2} - \left(\frac{W+1}{2}\log_2 \frac{W+1}{2} - \frac{W-1}{2}\log_2 \frac{W-1}{2}\right). \tag{100}$$

# 6 Security of AMQD for Gaussian Collective Attacks

The main result regarding the security of AMQD is summarized in Theorem 4. It states that under an optimal Gaussian collective attack, any crosstalk among the sub-channels cannot allow leaking of more information to an eavesdropper than the single-carrier CVQKD.

**Theorem 4** (Unconditional security of AMQD in presence of crosstalk, optimal Gaussian attack). *Let $\tilde{\gamma}(\mathcal{N}) > 0$ be a nonzero crosstalk among the sub-channels in an AMQD setting, and $\chi(B:E)_{single}$ be Eve's Holevo information under an optimal Gaussian collective attack for a single-carrier CVQKD. Then, $\chi(B:E)_{\tilde{\gamma}} \leq \chi(B:E)_{single}$, where $\chi(B:E)_{\tilde{\gamma}}$ is the Holevo information of Eve in AMQD modulation in the presence of nonzero crosstalk, under an optimal Gaussian collective attack.*

*Proof.*
The proof will assume reverse reconciliation, hence Eve's correlation will be quantified by the preserved correlation with Bob's system ($B$).

First, we express the eavesdropper information under an optimal Gaussian attack in an AMQD modulation. Then, we show that this attack – besides the fact that it maximizes Eve's Holevo information – cannot leak more information to an eavesdropper than the single-carrier case. (The optimality of Gaussian collective attacks for single-carrier CVQKD has already been proven in [20], [21] and the proof follows for the multicarrier case of AMQD. Therefore, we focus mainly on the analysis of the amount of leakable information in a multicarrier CVQKD scenario, assuming an optimal Gaussian collective attack.)

The notations used in our proof are summarized as follows. Let $\rho \in \mathcal{H}$ be the density matrix of a Gaussian state, and let $\rho' \in \mathcal{H}$ be the density matrix of a non-Gaussian state with the same covariance matrix and displacement vector. Let $p(x)$ refer to the Gaussian probability distribution, and $p'(x)$ to an arbitrary probability distribution with the same first and second momenta as $p(x)$. Let $\rho_{AB}$ be a Gaussian state between Alice and Bob, and let $\rho'_{AB}$ refer to an arbitrary shared state. For these density matrices, the relation between the joint entropy is $H(\rho_{AB}) \geq H(\rho'_{AB})$.

Let's assume that a Gaussian density matrix $\rho_i$ is transmitted through sub-channel $\mathcal{N}_i$, and the shared state between Alice, Bob and Eve that is related to the $i$-th sub-channel is a pure system $\rho_i^{ABE}$. The measurement process is characterized as follows. Let $X_i$ be Alice's classical variable that is encoded into the input system $A_i$, which is a density matrix $\rho_i$, and let $B_i$ be the classical variable of Bob that results from a measurement $M_i = \{M_{i,B_i}\}_{B_i}$ applied on $\rho_i$, where $M_{i,B_i}$ are positive operators, $\sum_i M_{i,B_i} M_{i,B_i}^\dagger = I$. The conditional entropy between $A_i$ and $B_i$ is $H(A_i|B_i) = H(\bar{A}_i) - H(B_i)$, where $\bar{A}_i$ refers to the measured density matrix $\bar{\rho}_i$ [20]. Let's denote Eve's optimal Gaussian attack on the $i$-th sub-channel $\mathcal{N}_i$ by the CPTP map $\mathcal{T}_i$. Taking arbitrary density matrices $\rho$ and $\sigma$, the relation $\mathcal{D}(\rho\|\sigma) \geq \mathcal{D}(\mathcal{T}_i(\rho)\|\mathcal{T}_i(\sigma))$ follows, where $\mathcal{D}(\cdot\|\cdot)$ is the relative entropy function, along with the entropic relation $H(A) - H(A') \geq H\mathcal{T}(A) - H\mathcal{T}(A')$ [20].

In the next step, we express the Holevo information of the eavesdropper $\chi(B_i : E_i)_{\gamma,\mathcal{N}_i}$ on a sub-channel $\mathcal{N}_i$, assuming reverse reconciliation, and optimal Gaussian attack, where $B_i$ is Bob's variable from the $i$-th sub-channel, while $E_i$ identifies Eve's variable. During the optimal Gaussian attack against AMQD, Eve attacks each sub-channel separately with a beam splitter that has transmittance $T_{Eve,i}$.

Assuming an average modulation variance $\sigma_\omega^2 = \frac{1}{l}\sum_l \sigma_{\omega_i}^2 < \sigma_{\omega_0}^2$ in the AMQD setting with $l$ sub-channels for the transmittance ($\sigma_{\omega_i}^2$-s are constant values due to the optimality consumption, see Theorem 2), the following relation holds for the Holevo information of the eavesdropper:

$$\begin{aligned}\tilde{\chi}(B:E) &= H(E) - H(E|B) \\ &= \frac{1}{l}\sum_l \left(H(E_i) - H(E_i|B_i)\right) \\ &\leq \chi(B:E)_{single} = H(E) - H(E|B)_{single}.\end{aligned} \quad (101)$$

This relation follows from that in an AMQD modulation $H(E|X) \geq H(E|X)_{single}$, because the same amount of information is transmitted via modulation variance $\sigma_\omega^2 < \sigma_{\omega_0}^2$ over the $l$ sub-channels.

At this point, we add the $\gamma > 0$ *crosstalk* into the picture. Assuming the use of sub-channel $\mathcal{N}_i$, the optimal Gaussian attack leads to

$$\begin{aligned}\chi(B:E)_{\gamma,\mathcal{N}_i} &= \chi(B:E)_{\mathcal{N}_i} + |T_{Eve,i}|^2 \gamma(\mathcal{N}_i) \\ &= \chi(B:E)_{\mathcal{N}_i} + \sum_{j\neq i}|T_{Eve,i}|^2 x_j \chi(A_j : B_j).\end{aligned} \quad (102)$$

The average modulation $\omega$ on the $l$ sub-channels leads to average Holevo information

$$\tilde{\chi}(B:E) = \frac{1}{l}\sum_l \chi(B:E)_{\mathcal{N}_i}, \quad (103)$$

while the average amount of crosstalk over all the $n$ sub-channels is evaluated as

$$\begin{aligned}\tilde{\gamma}(\mathcal{N}) &= \frac{1}{n}\sum_n \gamma(\mathcal{N}_i) \\ &= \frac{1}{n}\sum_n \sum_{j\neq i} x_j \chi(A_j : B_j),\end{aligned} \quad (104)$$

which together leads to the average Holevo information $\tilde{\chi}(B:E)_\gamma$ over all the $(n)$ sub-channels, as

$$\begin{aligned}
\tilde{\chi}(B:E)_{\tilde{\gamma}} &= \tfrac{1}{l}\sum_{l}\chi(B:E)_{\mathcal{N}_i} + \tfrac{1}{n}\sum_{n}|T_{Eve,i}|^2 \gamma(\mathcal{N}_i) \\
&= \tfrac{1}{l}\sum_{l}\chi(B:E)_{\mathcal{N}_i} + \tfrac{1}{n}\sum_{n}\sum_{j\neq i}|T_{Eve,i}|^2 x_j \chi(A_j:B_j) \\
&= \tfrac{1}{l}\sum_{l}\chi(B:E)_{\mathcal{N}_i} + \tfrac{1}{n}\sum_{j\neq i}|T_{Eve}|^2 x_j \chi(A_j:B_j) \\
&= \tfrac{1}{l}\sum_{l}\chi(B:E)_{\mathcal{N}_i} + \tfrac{1}{n}\sum_{n}|T_{Eve}|^2 \gamma(\mathcal{N}_i) \\
&= \tilde{\chi}(B:E) + |T_{Eve}|^2 \tilde{\gamma}(\mathcal{N}),
\end{aligned} \qquad (105)$$

where $|T_{Eve}|^2 = \tfrac{1}{n}\sum_n |T_{Eve,i}|^2$.

For this quantity, the relation

$$\tilde{\chi}(B:E)_{\tilde{\gamma}} \leq \chi(B:E)_{single} \qquad (106)$$

follows, since for $\tilde{\gamma}(\mathcal{N}) > 0$

$$\begin{aligned}
\tilde{\chi}(B:E) &= H(E) - H(E|B) \\
&< \tilde{\chi}(B:E)_{\tilde{\gamma}} = H(E) - H(E|B)_{\tilde{\gamma}} \\
&\leq \chi(B:E)_{single} = H(E) - H(E|B)_{single},
\end{aligned} \qquad (107)$$

and

$$H(E|B) > H(E|B)_{\tilde{\gamma}} \geq H(E|B)_{single}. \qquad (108)$$

The relation $H(E|B) > H(E|B)_{\tilde{\gamma}}$ holds for any $\tilde{\gamma}(\mathcal{N}) > 0$. The RHS of (108) is verified by the fact that a nonzero crosstalk has no relevance on the mutual information of Alice and Bob, i.e., communicating at a constant modulation variance $\sigma_\omega^2 < \sigma_{\omega_0}^2$ on the $l$ sub-channels, the relation $H(B|A) = H(B|A)_{\tilde{\gamma}} = H(B|A)_{single}$ follows for an AMQD modulation. The first equality follows from the fact that the presence of any nonzero crosstalk does not change on the mutual information of the legally parties, while the second comes from that the in an AMQD modulation the same amount information transmitted than single-carrier CVQKD via lower modulation variance (see (75)), from which (108) is immediately concluded.

Finally, we show that the Gaussian collective attack is optimal in the AMQD modulation. Fixing the first and second moments, under an optimal Gaussian attack for all sub-channels $\mathcal{N}_i$, $i = 1...n$, and with a nonzero crosstalk $\tilde{\gamma}(\mathcal{N}) > 0$, by the optimality consumption one obtains nonnegative $\Delta \geq 0$ for the difference of Eve's Holevo information that results from an optimal Gaussian and an arbitrary attack:

$$\begin{aligned}
\Delta &= \tilde{\chi}(B:E)_{\tilde{\gamma}} - \tilde{\chi}(B':E')_{\tilde{\gamma}} \\
&= \tfrac{1}{l}\big(\sum_l (H(E_i) - H(E'_i)) - \sum_l (H(E_i|B_i) - H(E'_i|B'_i))\big) \\
&\quad + \tfrac{1}{n}\sum_n \big(\sum_{j\neq i} |T_{Eve,i}|^2 x_j \chi(A_j:B_j) - \sum_{j\neq i} |T_{Eve,i}|^2 x_j \chi'(A_j:B_j)\big) \\
&= (H(E) - H(E')) - (H(E|B) - H(E'|B')) \\
&\quad + \big(|T_{Eve}|^2 \tilde{\gamma}(\mathcal{N}) - |T_{Eve}|^2 \tilde{\gamma}'(\mathcal{N})\big) \\
&= (H(AB) - H(A'B')) - (H(AB|X) - H(A'B'|X')) \\
&\quad + \big(|T_{Eve}|^2 \tilde{\gamma}(\mathcal{N}) - |T_{Eve}|^2 \tilde{\gamma}'(\mathcal{N})\big) \\
&= (H(AB) - H(A'B')) - (H(\overline{A}\overline{B}) - H(\overline{A}'\overline{B}')) + (H(X) - H(X')) \\
&\quad + \big(|T_{Eve}|^2 \tilde{\gamma}(\mathcal{N}) - |T_{Eve}|^2 \tilde{\gamma}'(\mathcal{N})\big),
\end{aligned} \qquad (109)$$

where $\tilde{\chi}(B':E')_{\tilde{\gamma}}$ is Eve's average Holevo information in an AMQD modulation for an arbitrary attack, $\tilde{\chi}(B:E)_{\tilde{\gamma}}$ is evaluated by (105), while $|T_{Eve}|^2 \tilde{\gamma}(\mathcal{N})$ and $|T_{Eve}|^2 \tilde{\gamma}'(\mathcal{N})$ refer to the leaked valuable crosstalk information under an optimal Gaussian and an arbitrary attack, respectively.

Since any nonzero crosstalk $\tilde{\gamma}(\mathcal{N}) > 0$ acts as a Gaussian noise for the receiver that is already included in the sub-channel noise variance $\sigma^2_{\mathcal{N}_i}$, see (64), the next equation is straightforward for the minimized mutual information under an optimal Gaussian collective attack:

$$\begin{aligned}
I(A:B) - I(A':B') &= \tfrac{1}{l}\big(\sum_l I(A_j:B_j) - \sum_l I(A'_j:B'_j)\big) \\
&= \tfrac{1}{l}\big(\sum_l (H(A_j) - H(A'_j)) + \sum_l (H(B_j) - H(B'_j)) \\
&\quad - \sum_l (H(A_j B_j) - H(A'_j B'_j))\big) \\
&= \tfrac{1}{l}\big(\sum_l (H(B_j) - H(B'_j)) - \sum_l (H(A_j B_j) - H(A'_j B'_j))\big) \\
&\leq 0.
\end{aligned} \qquad (110)$$

Thus, Eve's Gaussian collective attack remains optimal in an AMQD modulation (i.e., Eve's Holevo information is maximized), and the presence of a nonzero crosstalk has no effect on the mutual information between Alice and Bob.

On the other hand, Eve's Holevo information could be slightly improved by a nonzero crosstalk, to be precise by $|T_{Eve}|^2 \tilde{\gamma}(\mathcal{N})$, however an AMQD modulation does not allow leaking of more information to Eve than the single-carrier CVQKD. The argumentation behind this is as follows. The inequality

$$\begin{aligned}
\chi(B:E)_{single} - \tilde{\chi}(B:E)_{\tilde{\gamma}} &\geq 0 \\
\chi(B:E)_{single} - \big(\tilde{\chi}(B:E) + |T_{Eve}|^2 \tilde{\gamma}(\mathcal{N})\big) &\geq 0
\end{aligned} \qquad (111)$$

directly comes from (105), which combined with (108) leads to

$$\chi(B:E)_{single} - \tilde{\chi}(B:E) \geq |T_{Eve}|^2 \tilde{\gamma}(\mathcal{N}), \qquad (112)$$

which reveals that in an AMQD modulation, the degradation of Eve's Holevo information in comparison to the single-carrier CVQKD is always equal to or greater than the valuable information that is leaked from the crosstalk to Eve, thus

$$\tilde{\chi}(B:E) < \tilde{\chi}(B:E)_{\tilde{\gamma}} < \chi(B:E)_{single} - |T_{Eve}|^2 \tilde{\gamma}(\mathcal{N}), \tag{113}$$

hence for any $|T_{Eve}|^2 \tilde{\gamma}(\mathcal{N})$, the inequality

$$\chi(B:E)_{\tilde{\gamma}} \leq \chi(B:E)_{single} \tag{114}$$

immediately follows.

These results lead to the following key rates for reverse reconciliation at an AMQD modulation, under an optimal Gaussian collective attack:

$$R = I(A:B) - \tilde{\chi}(B:E)_{\tilde{\gamma}}, \tag{115}$$

and, if Bob is allowed to perform a collective measurement, then the rate is

$$R = \chi(A:B) - \tilde{\chi}(B:E)_{\tilde{\gamma}}. \tag{116}$$

These results prove the unconditional security of AMQD against optimal Gaussian collective attacks, which concludes the proof.

∎

# 7 Conclusions

The CVQKD protocols represent one of the most capable practical manifestations of quantum information theory. While the DVQKD protocols cannot be implemented within the framework of current technology, the CVQKD schemes can be established over standard communication networks and practical devices. Besides the attractive properties, the CVQKD schemes have an extreme sensitivity to the channel noise and other loss which allow no to use these protocols with such a high efficiency as it is available for traditional protocols in a traditional telecommunication scenario. To resolve the problem of low tolerable loss and excess noise, we introduced a new modulation scheme for CVQKD. The input Gaussian variables are transformed into several Gaussian subcarrier CVs, which are then transformed back by the continuous unitary CVQFT operation at the receiver. The transmission is realized through several Gaussian sub-channels, each dedicated to a given subcarrier with an independent noise variance. The AMQD modulation allows higher tolerable loss and excess noise in comparison with the standard modulation, and can be applied in both one-way and two-way CVQKD. We also investigated an adaptive modulation variance allocation mechanism for the scheme, which can significantly improve the efficiency of the transmission, particularly in the low SNR regimes.

# Acknowledgements

The author would like to thank Professor Sandor Imre for useful discussions. The results are supported by the grant COST Action MP1006.

# Supplemental Information

## S.1 Notations

The notations of the manuscript are summarized in Table S.1.

**Table S.1.** The summary of the notations.

| Notation | Description |
|---|---|
| $F(\cdot) = \mathrm{CVQFT}(\cdot)$ | The CVQFT transformation, applied on Bob's side, continuous variable $U$ unitary operation. |
| $F^{-1}(\cdot) = \mathrm{IFFT}(\cdot)$ | Inverse FFT transform, applied on Alice's side. |
| $\sigma^2_{\omega_0}$ | Single-carrier modulation variance. |
| $\sigma^2_\omega = \frac{1}{l}\sum_l \sigma^2_{\omega_i}$ | Multicarrier modulation variance. Average modulation variance of the $l$ Gaussian sub-channels $\mathcal{N}_i$. |
| $\mathbf{z} = \mathbf{x} + i\mathbf{p} = (z_1,\ldots,z_n)^T \in \mathcal{CN}(0,\mathbf{K_z})$ | An $n$-dimensional, zero-mean, circular symmetric complex random Gaussian vector, $\mathbf{K_z} = \mathbb{E}[\mathbf{zz}^\dagger]$, where $z_i = x_i + ip_i$, $\mathbf{x} = (x_1,\ldots,x_n)^T$ and $\mathbf{p} = (p_1,\ldots,p_n)^T$, and $x_i \in \mathbb{N}(0,\sigma^2_{\omega_0})$, $p_i \in \mathbb{N}(0,\sigma^2_{\omega_0})$ are i.i.d. zero-mean Gaussian random variables. |
| $\mathbf{d} = F^{-1}(\mathbf{z}) \in \mathcal{CN}(0,\mathbf{K_d})$ | An $n$-dimensional, zero-mean, circular symmetric complex random Gaussian vector, $\mathbf{K_d} = \mathbb{E}[\mathbf{dd}^\dagger]$, $\mathbf{d} = (d_1,\ldots,d_n)^T$, $d_i = x_i + ip_i$, $x_i \in \mathbb{N}(0,\sigma^2_{\omega_F})$, $p_i \in \mathbb{N}(0,\sigma^2_{\omega_F})$ are i.i.d. zero-mean Gaussian random variables, $\sigma^2_{\omega_F} = 1/\sigma^2_{\omega_0}$. The $i$-th component is $d_i \in \mathcal{CN}(0,\sigma^2_{d_i})$, with variance $\sigma^2_{d_i} = \mathbb{E}[|d_i|^2]$. |
| $|\phi_i\rangle = |F^{-1}(z_i)\rangle = |d_i\rangle$ | The $i$-th subcarrier Gaussian CV, where $F^{-1}$ stands for the inverse FFT. |
| $|\varphi_i\rangle = F(|\phi_i\rangle) = |F(F^{-1}(z_i))\rangle = |z_i\rangle$ | Bob's decoded Gaussian state, the CVQFT- |

| | |
|---|---|
| | transformed $\lvert\phi_i\rangle$ subcarrier CV. |
| $\mathcal{N}$ | Gaussian quantum channel in the single-carrier transmission. |
| $\mathcal{N}_i, i = 1,\ldots,n$ | Gaussian sub-channels in the multicarrier transmission. |
| $T(\mathcal{N}) = \operatorname{Re} T(\mathcal{N}) + \mathrm{i}\operatorname{Im} T(\mathcal{N}) \in \mathcal{C}$ | Channel transmittance, normalized complex variable. The real part identifies the position quadrature transmission, the imaginary part stand for the transmittance of the position quadrature. |
| $T(\mathcal{N}_i) = \operatorname{Re} T(\mathcal{N}_i) + \mathrm{i}\operatorname{Im} T(\mathcal{N}_i) \in \mathcal{C}$ | Transmittance of the $i$-th sub-channel. |
| $T_{Eve}$ | Eve's transmittance, $T_{Eve} = 1 - T(\mathcal{N})$. |
| $T_{Eve,i}$ | Eve's transmittance for the $i$-th subcarrier CV. |
| $W$ | Variance of Eve's EPR ancilla used in the entangling cloner attack. |
| $\Delta \in \mathcal{CN}(0,\mathfrak{C}(\Delta))$ | Gaussian noise of the quantum channel $\mathcal{N}$, zero-mean, circular symmetric complex Gaussian random vector with variance $\mathfrak{C}(\Delta) = \mathbb{E}[\Delta\Delta^{\dagger}]$, and with quadrature components $\Delta_{x_i} \in \mathbb{N}(0,\sigma^2_{\mathcal{N}_i})$, $\Delta_{p_i} \in \mathbb{N}(0,\sigma^2_{\mathcal{N}_i})$. |
| $\lvert\varphi'_i\rangle = \lvert x'_i + \mathrm{i}p'_i\rangle = \lvert z'_i\rangle$ | Bob's Gaussian coherent state in the phase space. Modeled as a zero-mean, circular symmetric complex Gaussian random variable $z_i \in \mathcal{CN}(0,\sigma^2_{z_i})$, $\sigma^2_{z_i} = \mathbb{E}[\lvert z_i\rvert^2]$, $z_i = x_i + \mathrm{i}p_i$, with quadrature components $x_i \in \mathbb{N}(0,\sigma^2_{\omega_0})$, $p_i \in \mathbb{N}(0,\sigma^2_{\omega_0})$, which are i.i.d. zero-mean Gaussian random variables. |
| $F(\mathbf{T}(\mathcal{N}))$ | The CVQFT transform of transmittance matrix $\mathbf{T}(\mathcal{N}) \in \mathcal{C}^n$, $n$ dimensional complex vector. |
| $F(\Delta) \in \mathcal{CN}(0,\mathfrak{C}(F(\Delta)))$ | The CVQFT transform of vector $\Delta$. An $n$-dimensional zero-mean, circular symmetric complex Gaussian random vector, where $\mathfrak{C}(F(\Delta)) = \mathbb{E}[F(\Delta)F(\Delta)^{\dagger}]$, and the quadrature components are $F(\Delta_{x_i}) \in \mathbb{N}(0,\sigma^2_{F(\mathcal{N}_i)})$, $F(\Delta_{p_i}) \in \mathbb{N}(0,\sigma^2_{F(\mathcal{N}_i)})$, where $\sigma^2_{F(\mathcal{N})} < \sigma^2_{\mathcal{N}}$. |
| $\mathbf{y}[j] = F(\mathbf{T}(\mathcal{N}))F(\mathbf{d})[j] + F(\Delta)[j],$ $j = 1,\ldots,n.$ | An AMQD block. Formulated by $n$ Gaussian subcarrier continuous variables, where $j$ is the index of the AMQD block, and $\mathbf{y}[j] = (y_1[j],\ldots,y_n[j])^T$ $F(\mathbf{d})[j] = (F(d_1)[j],\ldots,F(d_n)[j])^T,$ $F(\Delta)[j] = (F(\Delta_1)[j],\ldots,F(\Delta_n)[j])^T.$ |
| $\tau = \|F(\mathbf{d})[j]\|^2$ | Exponentially distributed variable, $\mathbb{E}[\tau] \leq n2\sigma_\omega^2$. |
| $\nu_i = \sigma_\mathcal{N}^2 \big/ |F(T_i)|^2$ | Ratio of noise variance and Fourier transformed channel transmittance, where $\left|F(T_i)\right|^2 = \left|\sum_{k=1}^n T_k e^{\frac{-i2\pi ik}{n}}\right|^2$, $i = 1\ldots n$. |
| $\nu_{Eve} = 1/\lambda$ | Security bound of the optimal Gaussian attack, where $\lambda$ is the Lagrange coefficient. |
| $\Omega = \Pr\left(\sum_l |F(T_i)|^2 = l \max_{\forall i} |F(T_i)|^2\right)$ | The probability that the sum of the squared magnitudes of the Fourier transformed coefficients of the $l$ Gaussian sub-channels picks up a maximum value. |
| $L = \left|F(T_{1\ldots n})\right|^2$ | Ordered list of the squared magnitudes so that $|F(T_i)|^2 \geq |F(T_{i+1})|^2$ and $\nu_i \leq \nu_{i+1}$. |
| $N_{single} = \frac{(W-1)\left(|T_{Eve}|^2\right)}{1-|T_{Eve}|^2}$ | Excess noise at a single-carrier transmission, where $T_{Eve}$ is Eve's transmittance, $W$ is the variance of the EPR-ancilla. |
| $N_{AMQD} = \frac{(W-1)\left(\frac{1}{n}\sum_n |F(T_{Eve,i})|^2\right)}{1-\frac{1}{n}\sum_n |F(T_{Eve,i})|^2}$ | The excess noise in the AMQD modulation scheme. |
| $\kappa = N_{single}/N_{AMQD} \geq 1$ | Ratio of single-carrier and AMQD excess noise. |
| $N_{tol,AMQD} = \alpha N_{tol,single}$ | Tolerable excess noise in AMQD modulation, where $\alpha = x\kappa \geq 1$. |
| $\eta_{single}, \eta_{AMQD}$ | Single-carrier and AMQD efficiency. |

## S.2 Abbreviations

| | |
|---|---|
| **AMQD** | Adaptive Multicarrier Quadrature Division |
| **AWGN** | Additive White Gaussian Noise |
| **BS** | Beam Splitter |
| **CV** | Continuous-Variable |
| **CVQFT** | Continuous-Variable Quantum Fourier Transform |
| DR | Direct-Reconciliation |
| DV | Discrete-Variable |
| FFT | Fast Fourier Transform |
| FWHM | Full Width at Half Maximum |
| IFFT | Inverse Fast Fourier Transform |
| OFDM | Orthogonal Frequency-Division Multiplexing |
| RR | Reverse Reconciliation |
| SNR | Signal-to-Noise Ratio |